\documentclass{IEEEojcsys_modified}

\usepackage{cite}
\usepackage{graphicx}
\usepackage{color,array}
\usepackage[ruled]{algorithm2e}
\usepackage[colorlinks,urlcolor=blue,linkcolor=blue,citecolor=blue]{hyperref}

\IEEEoverridecommandlockouts               
\usepackage[math,theorems]{kmacros}
\allowdisplaybreaks

\allowdisplaybreaks[3] 

\begin{document}


\title{Solving Decision-Dependent Games by Learning from Feedback} 

\editor{}

\author{Killian Wood\affilmark{1}}

\author{Ahmed Zamzam\affilmark{3}}

\author{Emiliano Dall'Anese\affilmark{1,2}}

\affil{Department of Applied Mathematics, University of Colorado Boulder, CO 80309 USA} 
\affil{Department of Electrical, Computer, and Energy Engineering, University of Colorado Boulder, CO 80309 USA} 
\affil{National Renewable Energy Laboratory, Golden, CO 80523 USA} 
\authornote{This work was supported by the National Science Foundation (NSF)  awards 1941896 and 2044946, and by the NSF Mathematical Sciences Graduate Internship Program.}

\markboth{Solving Decision-Dependent Games by Learning from Feedback}{K. Wood {\itshape ET AL}.}

\begin{abstract}
This paper tackles the problem of solving stochastic optimization problems with a decision-dependent distribution in the setting of stochastic strongly-monotone  games and when the distributional dependence is unknown. A two-stage approach is proposed, which initially involves estimating the distributional dependence on decision variables, and subsequently optimizing over the estimated distributional map. The paper presents guarantees for the approximation of the cost of each agent. Furthermore, a stochastic gradient-based algorithm is developed and analyzed for finding the Nash equilibrium in a distributed fashion. Numerical simulations are provided for a novel electric vehicle charging market formulation using real-world data.
\end{abstract}


\maketitle

\section{INTRODUCTION}
The efficacy of stochastic optimization~\cite{nemirovskistochastic} and stochastic games~\cite{lei2022stochastic,koshal2010single,liu2011stochastic,lei2020synchronous,franci2021stochastic} generally hinges on the premise that the underlying data distribution is stationary. This means that the distribution of the data, which parameterize the problem or the game, does not change throughout the execution of the algorithm used to solve the stochastic problem or game, and is neither influenced or dependent on time nor the optimization variables themselves. This is a common setup that has been considered when game-theoretic frameworks have been applied to problems in, for example, ride hailing~\cite{fabiani2022stochastic}, routing~\cite{bakhshayesh2021decentralized},  charging of electric vehicles (EVs)~\cite{fele2019scenario,paccagnan2018nash}, power markets~\cite{kannan2011strategic},  power systems~\cite{zhou2019online}, and in several approaches for training of neural networks~\cite{franci2021training}. However, this assumption can be invalid in a variety of setups in which the cost to be minimized is parameterized by data that is received from populations or a collection of automated control systems, whose response is uncertain and depends on the output of the optimization problem itself. As an example, in a competitive market for electric EV charging~\cite{fele2019scenario,tushar2012economics}, the operators seek to find the charging prices (i.e., the optimization variables) to maximize the revenue from EVs; however, the expected demand (i.e., the ``data'' of the problem) is indeed dependent on the price itself. More broadly, power consumption in power distribution grids depends on electric prices~\cite{mathieu2011examining}. A similar example pertains to ride hailing~\cite{fabiani2022stochastic}. 

To accommodate this scenario, the so-called stochastic optimization with \emph{decision-dependent distributions} (also known as \textit{performative prediction} \cite{perdomo2020performative}) posits that we represent the data distribution used in optimization instead as a \textit{distributional map} $x\mapsto D(x)$ where $x$ are decision variables~\cite{drusvyatskiy2022stochastic,miller2021outside,narang2022learning,perdomo2020performative,wood2023stochastic}. In this work, we study decision-dependent stochastic games in which players seeks to minimize their cost (based on their optimization variables) subject to other players optimization variables, and where the \emph{data distribution of each player depends on the actions of all players} (we will use the term player and agent interchangeably). 

We focus on solving the Nash equilibrium problem of a game, which is to find a decision from which no agent is incentivized by their own cost to deviate when played. Formally, the stochastic Nash equilibrium problem with decision-dependent distributions considered in this paper is to find a point $x^{*}=(x_{1}^{*},\hdots, x_{n}^{*})\in\real^{n}$ such that
\begin{align}
\label{eqn:nash_problem}
    x_{i}^{*} \in \argmin_{x_{i}\in\cX_i}  
    F_{i}(x_{i},x_{-i}^{*}), \quad \forall \, i \in\{ 1, \ldots, n\}
\end{align}
with $F_{i}(x_{i},x_{-i}^{*})$ defined as:
\begin{align}
F_{i}(x_{i},x_{-i}^{*}) : =
    \underset{z_{i}\sim D_{i}(x_{i},x_{-i}^*)}{\bbE}
    f_{i}(x_{i},x_{-i}^{*},z_{i})
\end{align}
where: $z_{i}$ denotes a random variable supported on $\real^{k_{i}}$, $f_{i}:\real^{d} \times \real^{k_{i}}\rightarrow \real$ is a scalar valued function that is convex and continuously differentiable in $x_{i}$, $\cX_i \subseteq \real^{d_{i}}$ is a compact convex set, and $D_{i}:\real^{d}\rightarrow \cP(\real^{k_i})$ is a \textit{distributional map} whose output is a probability distribution supported on $\real^{k_i}$.

 {
Standard stochastic first-order methods are insufficient for solving problems of this form. As we will demonstrate later in the paper, even estimating the expected gradient from samples requires knowledge of the probability density function associated with $D_{i}$\textemdash which is not possible in a majority of practical applications.
}

 Hereafter, we use the term ``system'' to refer to a population or a collection of automated controllers producing a response $z_i\in\real^{k_{i}}$ upon observing $x$.  {To illustrate our setup, consider again the example where each agent represents an EV charging provider. Here, $x_i\in\real^{d_{i}}$ represents the charging price at a station managed by provider $i$, expressed in \$/kWh. Correspondingly, $z_{i}$ indicates demand for the service at that price, while $f_{i}$ is the service cost (or the negative of the total profit) for provider $i$. This is an example of a competitive market in which the demand for service is a function of the price of all providers; see, for example, the game-theoretic approaches presented in~\cite{mediwaththe2018game,fele2019scenario} and the Stackelberg game presented in~\cite{tushar2012economics}. However, compared to existing game-theoretic models for EV markets, the framework proposed in this paper allows for an uncertain response of  EV owners to price variations; this randomness is difficult to model, as it it related to the drivers' preferences and other externalities such as the locations of the charging stations, etc.,  as explained in, e.g.,~\cite{mediwaththe2018game,latinopoulos2017response,daina2017electric}. }

Challenges in solving problems of this form typically stem from the that fact that the distributional maps $D_{i}$ are often unknown~\cite{sun2014bayesian,den2015dynamic,cheung2017dynamic,chen2021nonparametric}. To overcome this challenge, we propose a learning-based optimization procedure -- in the  spirit of the methods proposed for convex optimization in \cite{lin2023plug,miller2021outside} -- to tackle the multi-player decision-dependent stochastic game.  {The key idea behind this framework is that we first propose a parameterization for the distributional map in the system and estimate it from responses. Then, we use the estimated distributional map throughout the game without requiring further interaction with the system.}

\subsection{RELATED WORK} Our work incorporates themes from games, learning, as well as stochastic optimization with decision-dependent distributions. We highlight the relationship with this relevant literature below.

\emph{Games.} Within the context of games, our work is specifically focused on solving Nash equilibrium problems using gradient-based methods and a variational inequality (VI) framework. The literature on stochastic games is extensive; for a comprehensive yet concise review of the subject, we refer the reader to the tutorials  \cite{scutari2010convex} and~\cite{lei2022stochastic}; see also pertinent references therein. A common denominator of existing frameworks is that the  data distribution is stationary. The work of \cite{rosen1965existence} demonstrates that strictly monotone games have unique solutions and that gradient play converges to it. The modern approach of solving Nash equilibrium problems for continuous games via variational inequalities can be attributed to Facchinei and Pang \cite{facchinei2003finite,harker1990finite}. For solving strongly monotone variational inequalities, the projected gradient method is capable of converging linearly.

Our work adds the additional complexity of minimizing communication between agents and hence we use a distributed gradient approach in our optimization algorithm. Distributed gradient methods have been explored extensively in the literature on convex optimization, though less so in that of variational inequalities. We refer the reader to \cite{nedic2020distributed} for a review in the convex optimization setting, and \cite{kovalev2022optimal} for variational inequalities. 

\emph{Decision-Dependent Data}.  This paper contributes to the growing body of literature that studies stochastic optimization with decision-dependent data distributions. While the concept of decision-dependent distributions has existed for some time, its roots go back to the framework ``Performative Prediction'' and its use within the machine learning community \cite{perdomo2020performative}; this  work posits the formulation of optimization problems in which the data distribution is explicitly dependent on the optimization variables, and proposes repeated retraining (and the limit points thereof) as a solution. This is a family of points that solve the induced stationary optimization problem. 
For repeated retraining, convergence of various stochastic gradient algorithms are studied in \cite{drusvyatskiy2022stochastic,mendler2020stochastic} in the batch setting, and in the time-varying setting in \cite{cutler2021stochastic,wood2021online}. The extension of problems with  to games includes two-player zero-sum games in \cite{wood2021online},  {and general multiplayer games in \cite{narang2022learning}.
Additional recent extensions to this line of work include distributionally robust optimization\cite{basciftci2021distributionally,luo2020distributionally} and time varying optimization \cite{wood2021online,cutler2021stochastic}.}

Recent works have since pivoted towards directly solving the optimization problem by incorporating a model for the distributional map. In \cite{miller2021outside}, the authors provide conditions that ensure convexity of the expected cost, alongside a learning-based algorithm for finding solutions to problems with location scale families. The show that location scale families can be learned from the system prior to optimization without requiring further interaction with the system during optimization. The work of \cite{lin2023plug} extends this learning framework beyond location scale families and derives regret bounds that incorporate the associated approximation and statistical errors arising from the statistical learning problem. Outside of the distribution learning approach, derivative-free methods have been studied in \cite{miller2021outside,narang2022learning,wood2023stochastic}. While these methods do not depend on a learned model of the distributional map, they generally exhibit slower convergence rates. On the other hand, adaptive learning strategies, particularly for location-scale models, have shown promise, as discussed in \cite{narang2022learning} whereby the model is learned during optimization. However, their exploration remains limited to this specific model category.

This work complements the technical findings in \cite{lin2023plug} by offering a generalization of the so called ``plug-in'' approach presented in this work to the multi-agent setting. Relative to \cite{narang2022learning}, we focus on solving the Nash equilibrium problem for a general class of models by learning the distributional map prior to optimization\textemdash thereby reducing the required number of interaction with the system to receive a reasonable answer.

\subsection{CONTRIBUTIONS}
In this work, we provide the following contributions to the body of work on stochastic optimization and Nash equilibrium problems with decision-dependent distributions.
    \begin{itemize}
    \item[(i)]  {We propose an algorithm for finding a Nash equilibrium in stochastic games with decision-dependent distributions where: (i)  the distributional map for each player's cost is estimated from samples, and (ii) the estimated distributional map is used in gradient-based strategies.}

    \item[(ii)]  {We provide guarantees on the approximation error of distributional maps for a class of map learning problems.}

    \item[(iii)]  {We show that the parameterized cost approximates the ground-truth in high-probability.}

    \item[(iv)]  {We propose a stochastic gradient-based algorithm for solving a parameterized strongly-monotone  game, and we demonstrate linear convergence in expectation.}

    \item[(v)]  {Finally, we provide numerical simulations of an EV charging market formulation using real-world data. The EV market formulation is new in the context of energy markets, thus providing contributions in this area.}
    \end{itemize}

\subsection{ORGANIZATION}
In Section \ref{sec:two}, we provide necessary notation and background for our analysis. In Section \ref{sec:three} we discuss the proposed learning algorithm in detail and present our primary result. Section \ref{sec:four} discusses the details of the optimization stage. We provide our numerical simulations in Section \ref{sec:four}. Proofs of the results are provided in the Appendix. 

\section{Notation and Preliminaries}
\label{sec:two}
 {
Throughout the paper, $\real^{d}$ denotes the $d$-dimensional Euclidean space with inner product $\langle \cdot, \cdot \rangle $, and Euclidean norm $\norm{\cdot}$. For a matrix $X \in \mathbb{R}^{n \times m}$, $\norm{X}$ denotes the spectral norm.
For a given integer $n$, $[n]$ denotes the set $\{1, 2, \ldots, n\}$ and $\mathcal{S}^{n-1}$ denotes the Euclidean hypersphere in $n$ dimensions, $\{x\in\real^{n} \vert \ \norm{x}_{2}=1\}$. The symbol $\bfone_{d}$ is used to denote the $d$-dimensional vector of all ones. Given vectors $x\in \real^n$ and $z\in\real^m$, we let $(x,z) \in \real^{n+m}$ denote their concatenation.
}

 {
For a symmetric positive definite matrix $W\in\real^{d \times d}$, the weighted inner product is defined by $\langle x,y \rangle_{W} =  \langle x,Wy \rangle$ and corresponding weighted norm $\norm{x}_{W} = \sqrt{\langle x,x\rangle_{W}}$ for any $x,y\in\real^{d}$. The weighted projection onto a set $\cX\subseteq\real^{d}$ with respect to the symmetric positive definite matrix $W\in\real^{d \times d}$ is given by the map
\begin{equation}
    \proj_{\cX,W}(x) : = \argmin_{y\in\cX} \frac{1}{2}\norm{x-y}_{W}^{2}
\end{equation}
for any $x\in\real^{d}$.
}
\subsection{Probability measures}
 {
Throughout this work, we restrict our focus to random variables drawn from continuous probability distributions supported over the Euclidean space. When random variables $X,Y\in\real^{k}$ are equal in distribution, i.e., $P(X\leq x) = P(Y\leq x)$ for all $x\in\real^{k}$, we write $X \overset{d}{=}Y$.}

 {Our analysis includes study of sub-exponential random vectors. A univariate random variable $X\in\real$ is said to be sub-exponential with modulus $\theta>0$ provided that the survival function satisfies $\bbP(\vert X \vert \geq t) \leq 2\exp(-t/\theta)$ for all $t\geq 0$. By extension, a random vector $X\in\real^{k}$ is sub-exponential provided that $\langle u, X \rangle$ is a sub-exponential random variable for all $u\in\cS^{k-1}$.}

 {To compare probability distributions, we will be interested in computing the distance between their associated probability measures\textemdash for which we need a complete metric space. We let $\mathcal{P}(\real^{k})$ denote the set of finite first moment probability measures supported on $\real^{k}$ and write the Wasserstein-$1$ distance as
\begin{equation*}
    W_{1}(\mu,\nu) = \sup_{h\in \mathcal{L}_{1}} \bigg\{ 
    \bbE_{X\sim\mu}[h(X)] - \bbE_{Y\sim\mu}[h(Y)]
    \bigg\}
\end{equation*}
for any $\mu,\nu\in\mathcal{P}(\real^{k})$, where $\mathcal{L}_{1}$ is the set of all $1$-Lipschitz continuous functions $h:\real^{k}\rightarrow \real$. Under these conditions, the set $(\mathcal{P}(\real^{k}),W_{1})$ forms a complete metric space \cite{bogachev2012monge}.
}

\subsection{Games}
 {
We consider a game that consists of $n$ players. Each player has a cost function $F_{i}$, distributional map $D_{i}$, and decision set $\mathcal{X}_{i}\subseteq \real^{d_{i}}$. Hence, each player chooses a decision, or strategy $x_{i}\in\mathcal{X}_{i}\subseteq\real^{d_{i}}$. The concatenation of the decision variables is written as $x =(x_{1},\hdots,x_{n})\in\mathcal{X}\subseteq\real^{d}$ where $\mathcal{X} = \prod_{i=1}^{n} \mathcal{X}_{i}$ and $d=\sum_{i=1}^{n} d_{i}$. For a fixed agent $i$, we will decompose the decision $x$ as $x=(x_{i},x_{-i})$ where $x_{-i}\in\real^{d-d_{i}}$ is the strategy vector of all agents excluding the $i$th one. 
}

 {
The collection of costs $F_{i}$ and decision sets $\cX_{i}$ defines the game
\begin{equation}
\label{eqn:game}
    \min_{x_{i}\in\mathcal{X}_{i}} F_{i}(x_{i},x_{-i}) , \, \,\,\,\ i \in [n].
\end{equation}
A Nash equilibrium of this game is a point $x^{*}\in\mathcal{X}$ provided that 
\begin{equation}
    x^{*}_{i}\in\argmin_{x_{i}\in\cX_{i}} F_{i}(x_{i},x_{-i}^{*})
\end{equation}
for all $i\in[n]$. Intuitively, $x^{*}$ is a strategy such that no agent can be incentivized by its cost to deviate from $x_{i}^{*}$ when all other agents play $x_{-i}^{*}$. Finding Nash equilibria is the primary focus of this work.
}

 {
Games of this form are commonly cast into a variational inequality framework. This is due, in part, to the observation that the Nash equilibria $x^{*}\in\cX$ are the solutions to the variational inequality 
\begin{equation*}
    \langle x - x^{*}, G(x^{*}) \rangle \geq 0 , \,\,\,\, \forall \, x\in\cX ,  
\end{equation*}
where the gradient map $G:\real^{d} \rightarrow \real^{d}$ is defined as  
\begin{equation}
\label{eqn:vi_formulaion}
    G(x) = \left( \nabla_{1} F_{1}(x), \hdots, \nabla_{n} F_{n}(x)\right).
\end{equation}
Here, the notation $\nabla_{i}$ is used to represent the partial gradient $\nabla_{x_{i}}$. We will denote the set of Nash equilibrium of a game with gradient map $G$ and domain $\cX$ as $\texttt{Nash}(G,\cX)$. Existence of solutions to variational inequalities of this form is guaranteed provided that the set $\cX$ is convex and compact and the gradient map $G$ is monotone; uniqueness is guaranteed when $G$ is strongly-monotone \cite{facchinei2003finite}. We say that $G$ is $\alpha$-strongly-monotone on $\cX$ provided that there exists $\alpha>0$ such that 
\begin{equation}
    \langle x - y, G(x) - G(y) \rangle \geq \alpha\norm{x-y}^{2} , \,\,\, \forall \, x,y\in\cX, 
\end{equation}
and monotone when $\alpha=0$. In this work, we primarily focus on strongly-monotone games. While monotone games are tractable, methods for solving them with decision-dependent distributions require alternative gradient estimators\textemdash a topic we leave to future work.
}

\subsection{Monotonicity in Decision-Dependent Games}

 {
In this work, we introduce the additional complexity to the formulation in \eqref{eqn:game} that the $F_{i}$'s are the expected cost over a distributional map $D_{i}:\real^{d}\rightarrow \cP (\real^{k_{i}})$. In particular,   we write the cost as
\begin{equation}
F_{i}(x_{i},x_{-i}) : =
\underset{z_{i}\sim D_{i}(x_{i},x_{-i})}{\bbE}
f_{i}(x_{i},x_{-i},z_{i}).
\end{equation}
This can be written alternatively as the integral 
\begin{equation}
    F_{i}(x) = \int_{\real^{k_{i}}} f_{i}(x,z_{i}) p_{i}(z_{i},x) \text{d} z_{i}
\end{equation}
where $p_{i}$ is the probability density function for the distribution $D_{i}(x)$. When the integral satisfies the Dominated Convergence Theorem, computing the gradient amounts to differentiating under the integral and using the product rule. We then obtain 
\begin{equation}
\nabla_{i} F_{i}(x) = \hspace{-.2cm} \underset{z_{i}\sim D_{i}(x)}{\bbE}\left[ \nabla_{x_i}f_{i}(x,z_{i}) + f_{i}(x,z_{i})\nabla_{i}\log p_{i}(x;z_{i})
\right],
\end{equation}
where we recall that $G(x) = \left(\nabla_{1}F_{1}(x) , \hdots, \nabla_{n}F_{n}(x)\right)$.
In short, characterizing the gradient of this decision-dependent game requires assumptions not only on $f_{i}$, but also on the properties of the distributional map $D_{i}$. Sufficient conditions for strong monotonicity of the game in \eqref{eqn:game}
are due to \cite{narang2022learning} and are stated in terms of the \textit{decoupled} costs, given by
\begin{equation}
    F_{i}(x,y) = \underset{z_{i}\sim D_{i}(y)}{\bbE} f_{i}(x,z_{i})
\end{equation}
for all $x,y\in\real^{d}$, and their associated decoupled partial gradients 
\begin{equation}
    G_{i}(x,y) = \underset{z_{i}\sim D_{i}(y)}{\bbE} \nabla_{i} f_{i}(x,z_{i}),
\end{equation}
for all $x,y\in\real^{d}$ and 
\begin{equation}
    H_{i}(x,y) = \nabla_{y_{i}} \underset{z_{i}\sim D_{i}(y)}{\bbE}  f_{i}(x,z_{i})
\end{equation}
for all $x,y\in\real^{d}$. A key observation used in the proof is that $G_{i}(x) = \nabla_{i}F_{i}(x) = G_{i}(x,x) + H_{i}(x,x)$.
\begin{theorem}{(Strong Monotonicity, \cite{narang2022learning})}
\label{thm:sm}
Suppose that,
\begin{itemize}
    \item[(i)] For all $y\in\cX$, $x\mapsto G(x,y)$ is $\lambda$-strongly monotone,
    \item[(ii)] For all $x\in\cX$, $y\mapsto H(x,y)$ is monotone,
\end{itemize}
and that for all $i\in[n]$, 
\begin{itemize}
    \item[(iii)] For all $x\in\cX, z_{i}\mapsto \nabla_{i} f_{i}(x,z_{i})$ is $L_{i}$-Lipschitz continuous,
    \item[(iv)] $y\mapsto D_{i}(y)$ is $\gamma_{i}$-lipschitz continuous on $(\cP(\real^{k_{i}}),W_{1})$.
\end{itemize}
Set $\kappa = \sqrt{\sum_{i=1}^{n}(  \frac{\gamma_{i}L_{i}}{\lambda})^{2}}$. Then if $\kappa < 1/2$, $x\mapsto G(x)$ is $\alpha = (1-2\kappa)\lambda$-strongly monotone. \hfill $\Box$
\end{theorem}
}

\section{Learning-based Decision-Dependent Games}
\label{sec:three}
In this work, we aim to solve the stochastic Nash equilibrium problem with decision-dependent data distributions as formulated in~\eqref{eqn:nash_problem}. Methods for finding Nash equilibrium for games with decision dependent data distributions either use derivative free optimization, at the expense of an extremely slow rate, or use derivative information in conjunction with a learned model of the distributional map \cite{narang2022learning}.

\begin{figure}[!t]
    \centering
    \includegraphics[width=0.35\textwidth]{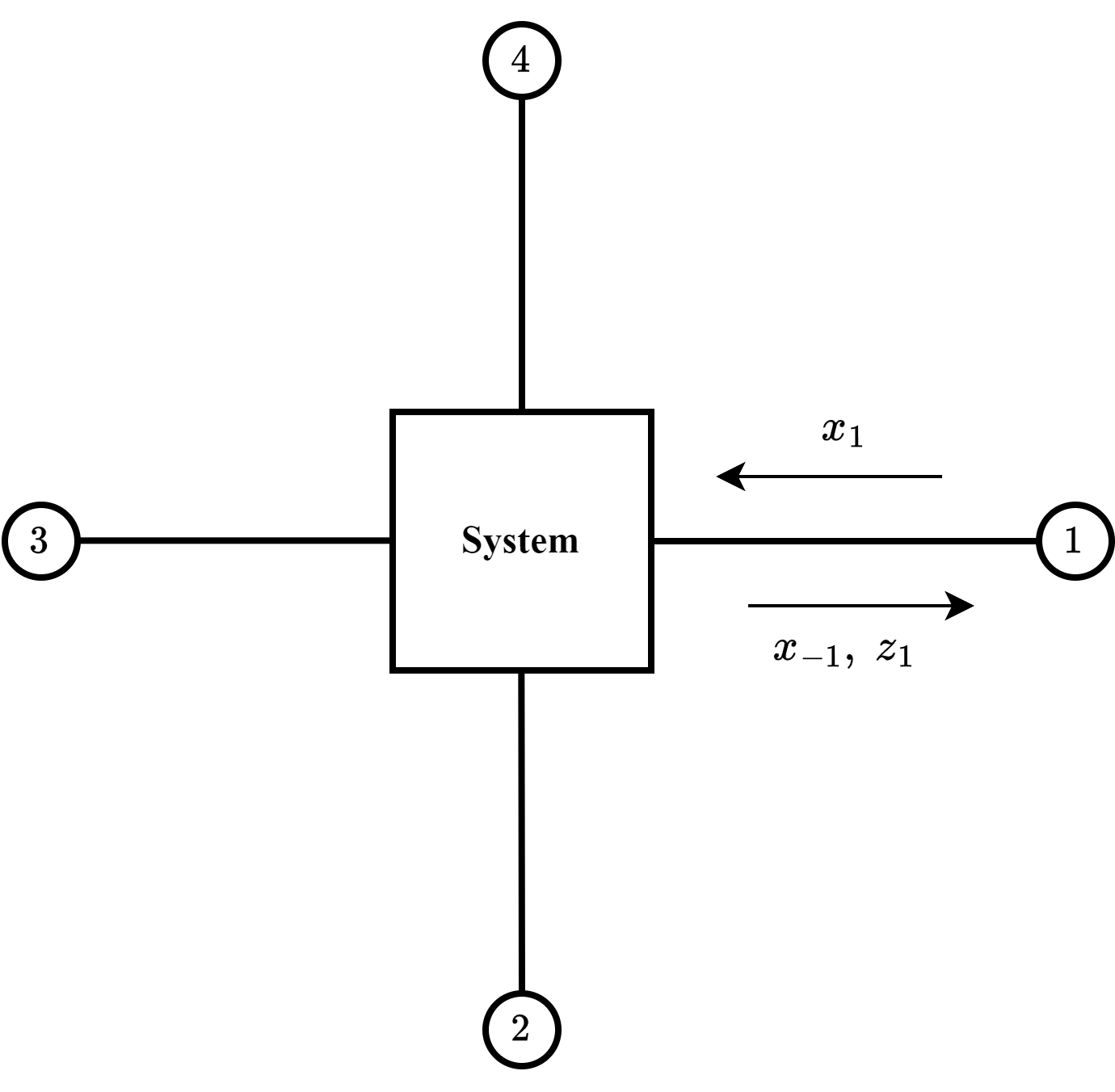}
    \caption{Communication structure allows agents to interact with the system in square by sending decision $x_{i}$. After deploying, agents can receive feedback from the system in the form of other agents decisions $x_{-i}$ and data $z_{i}$.}
    \label{fig:comms}
\end{figure}

In \cite{lin2023plug}, it is shown that a ``plug-in'' optimization approach, whereby a model for the distributional map is learned from samples prior to optimization, yields a bounded excess risk for the convex optimization problems with decision-dependent data.  {In this work, we leverage the properties of the system to simplify the communication structure of our approach, which we depict in Figure \ref{fig:comms}. We assume that realizations of $z_{i}$ can be directly observed from the system, and the decisions $x_{-i}$ can be obtained from a server or are made public (for example, the prices of EV charging  of different providers can be observed at the various stations). }

\begin{algorithm}[t]
\label{alg:algorithm}
\caption{Multi-phase Optimization}
 {
\KwIn{$m, \{D_{x_i}\}_{i=1}^{n}$}
\For{$j\in[m]$}{
    \For{$i\in[n]$}{
     Draw $x_{i}^{(j)}\sim D_{x_i}$ \;
    }
    Deploy $x^{(j)}$ \;
    Observe $z_{i}^{(j)}\sim D_{i}(x^{(j)})$ \;
}
\For{$i\in[n]$}{
    Fit $\hat{\beta}_{i} \in \argmin_{\beta_{i}\in\cB_{i}} \frac{1}{m}\sum_{j=1}^{m} R_{i}(x^{(j)}, z_{i}^{(j)}, \beta_{i}) $ \;
}
Compute $\hat{x} \in \texttt{Nash}(G_{\hat{\beta}}, \cX)$ \;
}
\end{algorithm}

 {To accommodate this setting, our algorithm proposes a multistage approach consisting of the following phases: (i)~sampling; (ii)~learning; (iii)~optimization. It is important to note that following the learning phase players only need to participate in gradient play without receiving any additional feedback from the system in the form of $z_{i}\sim D_{i}(x)$. This is distinct from existing approaches in which performatively stable points can only be reached after several (even thousands of) rounds of feedback \cite{perdomo2020performative, narang2022learning, wood2023stochastic}, and performatively optimal points can only be reached for models known to be location scale families a priori \cite{miller2021outside,narang2022learning}.}


\emph{Sampling.} In the sampling phase we require that players collaborate by each deploying a set of decisions $\{x_{i}^{(j)}\}_{j=1}^{m}\overset{i.i.d}{\sim} D_{x_i}$ so that they can collectively receive feedback $z_{i}^{(j)}\sim D_{i}(x^{(j)})$ from the system (in response to their deployed decisions $\{x_{i}^{(j)}\}_{j=1}^{m}$). The result is that each agent has access to a dateset $\{x^{(j)},z_{i}^{(j)}\}_{j=1}^{m}$ which they can use to learn their distributional map $D_{i}$

\emph{Learning.} In this procedure, each player will choose a hypothesis class of parameterized functions 
\begin{equation}
    \cH_{\cB_{i}} =\left\{ D_{\beta_{i}} \vert \ \beta_{i} \in \cB_{i} \subseteq \real^{ \ell_{i}}\right\},
\end{equation}
as well as a suitable criterion or risk function $R_{i}$, to formulate their own expected risk minimization problem 
\begin{equation}
\label{eqn:expected_risk}
    \beta^{*}_{i} \in \argmin_{\beta_{i} \in\cB_{i}} \underset{x\sim D_{x}, \\ z_{i}\sim D_{i}(x)}{\bbE}R_{i}(x,z_{i},\beta_i)
\end{equation}
over the random variable $(x,z_{i})$ drawn from the coupled distribution $(D_x, D_{i}(x))$.
Then, using the set of samples from the previous sampling phase, they can formulate the corresponding empirical risk minimization problem 
\begin{equation}
\label{eqn:erm}
    \hat{\beta}_{i} \in \argmin_{\beta_{i} \in\cB_{i}} \frac{1}{m}\sum_{j=1}^{m} R_{i}(x^{(j)},z_{i}^{(j)},\beta_i). 
\end{equation}
The result is a learned distributional map $D_{\hat{\beta}_{i}}$ approximating $D_{i}$, which we can now use to solve the approximate Nash equilibrium problem. 

\emph{Optimization.} Following the approximation phase, each player now has an learned model of their distributional map $D_{\hat{\beta}_{i}}$, which can be used to formulate an approximation of the ground-truth cost $F_{i}$ and hence an approximate Nash equilibrium problem:
\begin{align}
\label{eqn:approximate_nash}
    \hat{x}_{i} \in  \argmin_{x_{i}\in\cX_i} 
    F_{\hat{\beta}_{i}}(x_{i},\hat{x}_{-i}) 
\end{align}
for all $i\in[n]$, where 
\begin{align}
        F_{\hat{\beta}_{i}}(x_{i},\hat{x}_{-i}) := \hspace{-.2cm}
    \underset{z_{i}\sim D_{\hat{\beta}_{i}}(x_{i},\hat{x}_{-i})}{\bbE}
    f_{i}(x_{i},\hat{x}_{-i},z_{i}) \, .
\end{align}


\noindent  {Hereafter, we denote the Nash equilibrium of the approximate game as $\hat{x}$ to distinguish it from the ground truth $x^{*}$. In Algorithm 1, we write the set of Nash equilibria for the operator $G_{\hat{\beta}}$ with domain $\cX$ as $\texttt{Nash}(G_{\hat{\beta}},\cX)$. In practice, we will assume the appropriate assumptions to guarantee uniqueness of this assignment; in which case the set inclusion is simply an equality.}  

By solving \eqref{eqn:approximate_nash} instead of \eqref{eqn:nash_problem} we have introduced two errors: (i) the approximation error of the distributional map $D_{i}$ by elements of the hypothesis class $\cH_{\cB_i}$, and (ii) the estimation or statistical error by solving the ERM problem instead of the expected risk minimization problem. In \cite{lin2023plug}, the main result demonstrates that these two sources of error propagate through the optimization problem, and that the resulting excess risk can be bounded in terms of the sample complexity $m$. Our goal is to expand this result and provide additional analysis to our setting. 

\subsection{Parameter Estimation for Regular Problems}

A critical component of our analysis is the estimation or learning of the distributional map and the subsequent characterization of the estimation error. 
In this section, we outline a class of expected risk minimization problems, which we call \textit{regular problems}, for which we can characterize the distance between expected risk minimization solutions and empirical risk minimization solutions. Throughout, we write $R_{i}(\beta_{i}) = \bbE_{(x,z)}[R(x,z_{i},\beta_{i})]$ and $\hat{R}_{i}(\beta_{i}) = (1/m)\sum_{j=1}^{m}R_{i}(x^{(j)},z_{i}^{(j)},\beta_{i})$ for $\beta_{i}\in\real^{\ell_{i}}$ to denote the expected and empirical risk, respectively.

\begin{definition}{(Map Learning Regularity)} 
\label{as:regular}
A map learning problem, consisting of the optimization problems with costs $R_{i}$ and $\hat{R}_{i}$ over $\cB_{i}$, is regular provided that: 
\begin{itemize}
    \item[(a)] \textbf{Convexity}: The expected risk $\beta_i\mapsto R_{i}(\beta_i)$ is $\mu_{i}$-strongly convex, and the empirical risk $\beta_{i}\mapsto\hat{R}_{i}(\beta_{i})$ is convex. 
    \item[(b)] \textbf{Smoothness}: For all realizations of $x\in\cX$ and $z_{i}\in\real^{k_{i}}$, $\beta_{i} \mapsto \nabla_{\beta_{i}} R_{i}(x,z_{i},\beta_{i})$ is $L_{\beta_{i}}$-Lipschitz continuous.  
    \item[(c)] \textbf{Boundedness}:  {The set $\cB_{i}\subseteq\real^{\ell_{i}}$ is convex and compact.}
    \item[(d)]  \textbf{Sub-Exponential gradient}: For all $\beta_{i}\in\cB_{i}$, $\nabla_{\beta_{i}} R_{i}(x,z_{i},\beta_{i})$ is a sub-exponential vector with parameter $\theta_{i}>0$. \hfill $\Box$
\end{itemize}
\end{definition}

 {Items (a) and (c), taken together, guarantee existence of $\hat{\beta}$ and uniqueness of $\beta^{*}$ as defined in \eqref{eqn:erm} and \eqref{eqn:expected_risk}, respectively. Furthermore, the inclusion of item (b) is necessary to guarantee that first-order stochastic gradient methods will converge at least sub-linearly to $\hat{\beta}$. }
Lastly, the heavy-tail assumption~\cite{vershynin2018high} will allow us to describe the concentration of the gradient estimates. Together, they allow us to relate the solutions to the sample complexity in the following lemma.

\begin{lemma}{(Uniform Gradient Bound)}
\label{lemma:gradient_bound}
If the smoothness and sub-exponential gradient assumptions in Definition \ref{as:regular} hold for player $i \in [n]$, then for any $\delta\in(0,1/2)$ and any $m$ such that $m/\log(m) \geq 2(\ell_{i}+\log(1/\delta))$, we have that:
\begin{equation}
\sup_{\beta\in\cB} \norm{ \nabla \hat{R}_{i}(\beta) - \nabla R_{i}(\beta) } \leq C_i\sqrt{\frac{ \log(m)(\ell_{i} + \log(1/\delta))}{m}}
\end{equation}
with probability at least $1-\delta$, where $C_i = 4\max\{L_{\beta_{i}}/15r_{i}, \theta_{i}\}$. \hfill $\Box$
\end{lemma}
The proof of this result is provided in the Appendix~\ref{ap:lemmagr}. This result offers a broad generalization of  \cite[Equation (19b)]{mou2019diffusion} to any risk with Lipschitz-continuous sub-exponential gradients over any convex and compact set. Our result is comparable to the $\mathcal{O}(\sqrt{\ell_i}{m})$
rate that can be found for  specific problem instances such as linear least squares regression and logistic regression, but with the addition of a $\sqrt{\log m}$ factor. Indeed, the generality of the risk function requires that we enforce compactness of the domain, thus giving rise to this extra logarithmic factor. This gradient estimation result will now allow us to reach our desired bounded distance result, which we present in the following theorem.

\begin{theorem}{(ERM Approximation)}
\label{theorem:beta_estimation}
If the map learning problem is regular for  player $i \in [n]$ (i.e., it satisfies the assumptions in Definition~\ref{as:regular}), then for any $\delta\in(0,1/2)$ and any $m$ such that $m/\log(m) \geq 2(\ell_i+\log(1/\delta))$ we have that:
\begin{equation}
\label{eqn:beta_bound}
    \norm{\hat{\beta}_{i} - \beta_{i}^{*}}
    \leq C_i'\sqrt{\frac{ \log(m)(\ell_i + \log(1/\delta))}{m}}
\end{equation}
with probability at least $1-\delta$, where $C_i' = (4/\mu_{i})\max\{L_{\beta_i}/15r_i, \theta_i\}$. \hfill $\Box$
\end{theorem}
The proof of Theorem~\ref{theorem:beta_estimation} is provided in the Appendix~\ref{ap:beta_estimation}. The power in this characterization lies in the fact that it holds for any statistical learning problem satisfying the assumptions listed in Definition~\ref{as:regular}, and is not specific to the setting of learning distributional maps.
We note that our Definition~\ref{as:regular}, which is a property used in the Theorem~\ref{theorem:beta_estimation}, is different from the one in~\cite{lin2023plug} and it involves conditions that are easier to check.

As an example, we provide conditions for which a  linear least squares problem satisfies the regularity conditions and hence is subject to the above ERM approximation result.

\begin{proposition}{(Linear Least Squares Regularity)} 
\label{prop:ls}
Consider the linear least squares problem with expected risk problem
\begin{equation*}
    B_i^{*} \in \argmin_{B\in\cB_i} \frac{1}{2} \bbE_{(x,z)} \norm{Bx-z}^{2},
\end{equation*}
and empirical risk minimization problem 
\begin{equation*}
    \hat{B}_i \in \argmin_{B\in\cB_i} \frac{1}{2m}\sum_{j=1}^{m}\norm{Bx_i^{(j)}-{z_i^{(j)}}}^{2}.
\end{equation*}
Let $x_i\sim D_{i}$ with zero mean and covariance matrix $\Sigma_{i}$. If \\
(i) There exist $\gamma_{i},L_{i}>0$ such that $\gamma_{i} I \leq \Sigma_{i} \leq L_{i} I$,\\ 
(ii) The entries of $xx^{T}$ and $zx^{T}$ are sub-exponential,\\ 
(ii) The constraint set $\cB_{i}$ is convex and compact. \\ 
Then, the map learning problem is regular.  \hfill $\Box$
\end{proposition}

The proof of Proposition~\ref{prop:ls}
is provided in Appendix~\ref{ap:propls}. Deriving conditions for the more general case of non-linear regression is attainable but outside the scope of this work.  


\subsection{Bounding the Approximation Error}

Finding a relationship between $\hat{x}$ and $x^{*}$ will require that we first characterize an appropriate hypothesis class of distributions for learning. Here, we formalize the notion of misspecification and sensitivity for a hypothesis class $\cH_{\cB_{i}}$. 

\begin{definition}{(Misspecification~\cite{lin2023plug})} A hypothesis class $\cH_{\cB_{i}}$ is $\gamma_{i}$-misspecified provided that there exists a $\gamma_{i}>0$ such that 
\begin{equation}
    W_{1}(D_{\beta_{i}^{*}}(x), D_{i}(x)) \leq \eta_{i}
\end{equation}
for all $x\in\cX$. \hfill $\Box$
\end{definition}

Observe that if $\eta_i=0$, then our approximation error is zero and we incur no bias in our problem formulation. This corresponds to the case where the map $D_{i}$ is realizable \cite{shalev2014understanding}. Though, small values of $\eta_i$ are appropriate as $D_{i}$ may be complex enough that approximating it with a finite-dimensional parameter space is difficult. 

\begin{definition}{(Sensitivity~\cite{lin2023plug})}
\label{def:sensitive}
The hypothesis class $\cH_{\cB_{i}}$ is $\varepsilon_{i}$-sensitive if, for any $\beta_{i},\beta_{i}'\in\cB_{i}$,
\begin{equation}
    W_{1}(D_{\beta_{i}}(x), D_{\beta_{i}'}(x)) \leq \varepsilon_{i} \norm{\beta_{i}- \beta_{i}'}
\end{equation}
for all $x\in\cX$. \hfill $\Box$
\end{definition}


Sensitivity of $\cH_{\cB_{i}}$ is merely a convenient name for the condition that $\beta\mapsto D_{\beta_{i}}(x)$ be $\varepsilon_{i}$-lipschitz continuous for all realizations of $x\in\cX$. 
In the result that follows, we demonstrate that an appropriately misspecified and sensitive hypothesis class  {induces a cost that has bounded distance to the ground truth cost in \eqref{eqn:nash_problem}.} 

 {
\begin{theorem}{(Bounded Approximation)}\label{theorem:main} Suppose that the following conditions hold for all $i\in[n]$: \\
(i) The hypothesis class $\cH_{\cB_{i}}$ is $\eta_{i}$-misspecified, and $\varepsilon_{i}$-sensitive. \\
(ii) The map learning problem is regular.\\
(iii) For all $x\in\cX_i$, $z_{i}\mapsto f_{i}(x,z_{i})$ is $L_{z_i}$-Lipschitz continuous.  \\
Then, the bound
\begin{equation}
\label{eqn:infty_norm}
 \vert F_{\hat{\beta}_{i}}(x) - F_{i}(x) \vert 
\leq \eta_{i} L_{z_i}  +  L_{z_i} \varepsilon_{i} C_i' \zeta_i(m,\delta),
\end{equation}
holds with probability $1-\delta$ for any $x\in\cX$, where 
\begin{equation}
    \zeta_i(m,\delta) := \sqrt{\frac{ \log(m)(\ell_i + \log(1/\delta))}{m}} \, .
\end{equation}
and where $C_{i}'$ is as in $\eqref{eqn:beta_bound}$. \hfill $\Box$
\end{theorem}
} 

The proof of Theorem~\ref{theorem:main} is provided in the Appendix~\ref{ap:theorem_main}. By including an additional assumption on the Lipschitz continuity of the cost $f_{i}$  with respect to to other agents decisions $x_{-i}$, we can provide an analog of the excess risk result in ~\cite{lin2023plug}.
 {
\begin{corollary}
\label{cor:excess_risk}
Suppose that for all $i \in [n]$, the function $x_{-i} \mapsto F_{\beta_i}(x_i, x_{-i})$ is $L^\beta_{-i}$-Lipschitz on $\cX_{-i}$ for any $x_i \in \cX_i$. Then,
\begin{align}
\vert F_{i}(\hat{x}) - F_{i}(x^{*}) \vert \leq & ~2 \gamma_{i} L_{z_i}  + 2 L_{z_i} \varepsilon_{i}  C_i' \zeta_i(m,\delta) \nonumber \\
& + 2 \sqrt{2} \bar{L}_i \diam(\cX)
\label{eqn:excess_risk}
\end{align}
hold with probability $1-\delta$ for any $\hat{x}\in\texttt{NASH}(G_{\hat{\beta}},\cX)$ and $x^{*}\in\texttt{NASH}(G,\cX)$, $\bar{L}_i := \max\{L_{i}^{\beta^*}, L_{-i}^{\beta^*},L_{i}^{\hat{\beta}}, L_{-i}^{\hat{\beta}}\}$, and $C_{i}'$ is as in \eqref{eqn:beta_bound}. \hfill $\Box$
\end{corollary}
}

 {The analysis in this section demonstrates that the estimation procedure in  Algorithm \ref{alg:algorithm} yields a cost function that approximates the original cost in \eqref{eqn:nash_problem} with an error the decreases as the number of samples increases. Furthermore, this bound exists independent of the conditioning of the Nash equilibrium problem we solve in the optimization phase. We note that~\eqref{eqn:infty_norm} is similar to the result in~\cite{lin2023plug}, but it is based on a different definition of regular problem (see Definition~\ref{as:regular}); the bound~\eqref{eqn:excess_risk} is unique to this paper. }

In the section that follows, we examine a family of hypothesis classes that allows the approximated game to be monotone, and provide suitable algorithms for solving them with convergence guarantees.

\section{Solving Strongly-monotone Decision-dependent Games}
\label{sec:four}
 {
Since the agents lack full knowledge of the system and hence the ground truth distributional map $D_{i}$ in \eqref{eqn:nash_problem}, we cannot hope to enforce that $D_{i}$ satisfy any assumptions to encourage tractability of our optimization problem. We can however impose conditions on the hypothesis class  $\cH_{\cB_{i}}$, which is chosen by the agents. To successfully find a Nash equilibrium of the approximate problem in \eqref{eqn:approximate_nash}, it will be crucial that agents choose a class that balances expressiveness of the system (thereby making $\eta_{i}$ small) with tractability of the optimization. }

 {Perhaps the simplest model capable of achieving this goal is the location-scale family~\cite{miller2021outside,narang2022learning,wood2023stochastic}. In our setting, a location scale family parameterization for agent $i$ is a distributional map $D_{B_i}$ having matrix parameter $B_{i}\in\real^{k_i \times d}$ where $z_{i}\sim D_{B_i}$ if and only if 
\begin{equation}
    z_{i} \overset{d}{=} \xi_{i} + B_{i}x 
\end{equation}
for stationary random variable $\xi_i\sim D_{\xi_i}$. We note that this parameterization can be written alternatively as $z_{i} \overset{d}{=} \xi_i + B_{i}^{i}x_{i} + B_{-i}^{i}x_{-i}$, where $B_{i}^{i}\in\real^{k_{i}\times d_{i}}$ and $B_{-i}^{i}\in\real^{k_{i}\times(d-d_{i})}$ are block matrices such that $B_{i}x = B_{i}^{i}x_{i} + B_{-i}^{i}x_{-i}$ due to linearity. The resulting partial gradient has the form 
\begin{equation*}
    \nabla_{i}F_{i}(x) = \bbE_{z_{i}\sim D(x)}\left[\nabla_{i}f_{i}(x,z_{i}) + (B_{i}^{i})^{T}\nabla_{z_i}f_{i}(x,z_{i})\right], 
\end{equation*}
which is typically much simpler to analyze than alternative models.
Intuitively, this model allows us to express $z_{i}$ as the sum of a stationary random variable from a base distribution with a linear factor depending on $x$, where the matrix parameter $B_i$ weights the responsiveness of the population to the agents decisions. }

 {This model is particularly appealing as guarantees for learning $B_{i}$ are known and established in Proposition \ref{prop:ls}. Moreover, the matter of expressiveness is due to the fact that location scale families are a particular instance of strategic regression \cite{perdomo2020performative,lin2023plug}, in which member of the population interact with agents by modifying their stationary data (such as features in a learning task) $\xi_{i}$ in an optimal way upon observing $x$: 
\begin{equation*}
z_{i} \overset{d}{=} \argmin_{y} \left[ -u_{\beta_{i}}(x,y) + \frac{1}{2}\norm{y-\xi_{i}}^{2} \right],
\end{equation*}
where $u_{\beta_{i}}$ is a utility function parameterized by $\beta_{i}\in\cB_{i}$ corresponding to the utility that members of the population derive from changing their data in response to the decisions in $x$; and the quadratic term $1/2\norm{y-\xi_{i}}^{2}$ is the cost of changing their data from $\xi_{i}$ to $y$. Indeed when $u_{\beta_i}(x,) = \langle y,B_{i}x\rangle$ for $\beta_{i} = B_{i}\in\real^{k_i\times d}$, we recover the form above.}

 {Furthermore, location scale families immediately satisfy several of the assumption required for further analysis. In particular, it is known that Sensitivity (Definition \ref{def:sensitive}) holds with $\varepsilon_{i} = \max_{x\in\cX}\norm{x}^{2}$, Lipschitz continuity of $x\mapsto D_{B_{i}}$ holds with $\gamma_{i} = \norm{B_{i}}^{2}$, and Lipschitz continuity of $G_{B_{i}}$ holds due to the following result.}
 {
\begin{lemma}{(Lipschitz Gradient, \cite{narang2022learning})}
    Suppose that $D_{\beta_i}$ is such that  $z\overset{d}{=}B_{i}x + \xi_{i}$ with $\beta_{i} = B_{i}$, and that for each $i\in[n]$ there exists $\zeta_{i}\geq0$ such that $(x,z_{i})\mapsto \nabla_{i,z_{i}}f_{i}(x,z_{i})$ is $\zeta_{i}$-Lipschitz continuous. Then $G_{\beta_i}$ is $L$-Lipschitz continuous with 
    \begin{equation}
        L := \sqrt{\sum_{i=1}^{n} \zeta_{i}^{2}\max\{1,\norm{B_{i}^{i}}^{2}\}(1+\norm{B_{i}}^{2})
        } \, .
    \end{equation}
    \hfill $\Box$
\end{lemma}
}

 {Strong monotonicity will follow from Theorem \ref{thm:sm} provided that $G_{\beta_{i}}$ satisfy the remaining hypothesis on the $f_{\beta_i}$\textemdash which tends to be on a case-by-case basis. We will not require that $G_{\beta_i}$ use this parameterization in our analysis, however we can proceed with knowledge a model class satisfying our hypothesis does exist.}

\subsection{Distributed Gradient-based Method}
 {In our optimization phase, we seek to use a gradient-based algorithm that respects the agent's communication structure with the system. For the sake of readability, we will suppress the $\beta_{i}$ subscript and instead refer to quantities $G_{i}$ keeping in mind that they will correspond to the approximate Nash equilibrium problem in \eqref{eqn:approximate_nash} with solution $\hat{x}$.}

 {We will assume that each agent has access to an estimator of the gradient $\nabla_{i} F_{i}$ and is capable of projecting onto their decision set $\cX_{i}$. In the constant step-size setup, each agent chooses a rate $\omega_{i}>0$ and performs the update }

 {
\begin{equation*}
\label{eq:iteration}
    x_{i}^{t+1} = \proj_{\cX_{i}}  \left(x_{i}^{t} - \omega_{i}^{-1}g_{i}^{t}\right),
\end{equation*}
where $g_{i}^{t}$ is a stochastic gradient estimator for $\nabla_{i}F_{i}$ used at iteration $t$, which is then reported to the system and made available to all agents. For the sake of analysis, we will assume without loss of generality that the steps-sizes satisfy the ordering
\begin{equation*}
    \omega_{1} \geq \omega_{2} \geq ~\hdots~ \geq \omega_{n}
\end{equation*}
and hence $\omega_{1} = \max_{i\in[n]} \omega_{i}$ and $\omega_{n} = \min_{i\in[n]} \omega_{i}$. The collective update can be written compactly as 
\begin{equation}
\label{eq:weighted_iteration}
    x^{t+1} = \proj_{\cX,W}  \left(x^{t} - W^{-1}g^{t}\right),
\end{equation}
where $W = \diag(\omega_{1}\bfone_{d_1},\hdots,\omega_{n}\bfone_{d_n})$ and $g^{t}$ is an estimator for $G(x^{t})$ at iteration $t$. Convergence of this procedure hinges on the following assumptions. }

 {
\begin{assumption}
\label{ass:mono_lip}
The gradient function $G:\cX\subseteq\real^{d}\rightarrow \real^{d}$ is $\alpha$-strongly monotone and $L$-Lipschitz continuous. 
\end{assumption}
}

 {
\begin{assumption}(Stochastic Framework)
\label{ass:stochastic}
Let $\bbF=(\cF_{t})_{t\geq0}$ with elements
\begin{equation}
    \cF_{t} = \sigma(g^{\tau},\tau \leq t)
\end{equation}
be the natural filtration of the Borel $\sigma$-algebra over $\real^{d}$ with respect to $g^{t}$, and use the short-hand notation $\bbE_{t} \cdot  := \bbE_{z\sim D(x^{t})}[\cdot \vert \cF_{t}]$ as the conditional expectation over the product distribution $D(x^{t}) = \prod_{i=1}^{n} D_{i}(x^{t})$. There exist bounded sequences $\{\rho^{t}\}_{t\geq0},\{\sigma^{t}\}_{t\geq0}\subseteq\real_{+}$ such that 
\begin{align*}
     \text{(\textbf{Bias})} & \qquad \qquad \norm{\bbE_{t}g^{t}-G(x^{t})}\leq \rho^{t}\\ 
     \text{(\textbf{Variance})} & \qquad \qquad \bbE_{t}\norm{g^{t}-\bbE_{t}g^{t}}^{2}\leq (\sigma^{t})^{2}
\end{align*}
 where $\rho^{t}\leq \rho$ and $\sigma^{2}\leq \sigma$ for all $t\geq0$. 
\end{assumption}
}

 {
Assumption \ref{ass:mono_lip} is standard for guaranteeing convergence of gradient play \cite{facchinei2003finite}, and the uniformly bounded variance component of Assumption~\ref{ass:stochastic} is standard for convergence for stochastic algorithms. As we will show shortly, convergence with bias is possible and the result reduces to the unbiased case when $\rho^{t}=0$. The next result will quantify the one-step improvement of~\eqref{eq:iteration}.
}

 {
\begin{lemma}
{(One-step Improvement)} 
\label{lem:step_improvement}
Let Assumptions~\ref{ass:mono_lip} and \ref{ass:stochastic} hold. Then, the sequence generated by iteration~\eqref{eq:weighted_iteration} satisfies: 
\begin{align*}
    \bbE_{t}\norm{x^{t+1} - \hat{x}}_{W}^{2} 
    \leq 
     &\frac{\omega_{1}}{\omega_{n} + \alpha} \norm{x^{t}-\hat{x}}_{W}^{2}  \nonumber \\ 
     & + \frac{2\omega_{1}\left(\omega_{1}\rho^{2} + \alpha\sigma^{2}\right)}{\alpha\omega_{n}\left(\omega_{n}+\alpha\right)}
\end{align*}
for all $t \geq 0$, provided that $\omega_{1}/\omega_{n}^{2}\leq \alpha/(4L^{2})$. \hfill $\Box$
\end{lemma}
}

 {The proof of Lemma~\ref{lem:step_improvement} is provided in the Appendix~\ref{ap:step_improvement}. We note that setting $\omega_{i} =\omega$ for some $\omega>0$ recovers the result in \cite[Theorem 15]{narang2022learning}. Following this one-step analysis, we can show convergence to a neighborhood of the Nash equilibrium. }

 {
\begin{theorem}{(Neighborhood Convergence)}
\label{thm:nbhd_convergence}
Let Assumptions~\ref{ass:mono_lip} and \ref{ass:stochastic} hold, and suppose that $(\omega_{1} - \omega_{n})< \alpha$. Then, 
\begin{equation}
\limsup_{t\to\infty}\bbE\norm{x^{t}- \hat{x}}^{2}
    \leq \frac{2\omega_{1}\left(\omega_{1}\rho^{2} + \alpha\sigma^{2}\right)}{\alpha\omega_{n}\left(\omega_{n}+\alpha\right)}.
\end{equation}
\end{theorem}
}

 {The proof can be found in Appendix \ref{ap:nbhd_convergence}. The result shows that the algorithm converges linearly to a neighborhood of the Nash equilibrium $\hat{x}$, where the radius of the neighborhood is dictated by the step-size, variance, and bias bounds. When $\rho=\sigma=0$, we retrieve linear convergence. In order to converge to $\hat{x}$ directly, we will require a decaying step-size policy. For example, we consider the following policy:  
\begin{equation}
    \omega^{t} = \frac{\alpha(r+t-2)}{2}
\end{equation}
for fixed constant $r>2$, which we assumed to be shared by all agents. Hence, the decaying step-size update is given by 
\begin{equation}
\label{eq:iteration}
    x^{t+1} = \proj_{\cX}  \left(x^{t} - (\omega^{t})^{-1}g^{t}\right).
\end{equation}
In the theorem that follows, we show that this sequence converges to $\hat{x}$ provided that the bias shares an asymptotic rate with $(\omega^{t})^{-1}$.
}

 {
\begin{theorem}{(Convergence)}
\label{thm:convergence}
Suppose that Assumptions~\ref{ass:mono_lip} and \ref{ass:stochastic} hold and that there exists $\bar{\rho},s>0$ such that
\begin{equation}
    \norm{\bbE_{t}g^{t}-G(x^{t})}\leq \frac{\bar{\rho}}{s+t}
\end{equation}
for all $t\geq0$. Then,
\begin{equation}
    \bbE\norm{x^{t}-x^{*}}^{2}\leq \frac{A}{\alpha^{2}(r+t)}
\end{equation}
where 
\begin{equation*}
    A = \max \left\{ \alpha^{2}r\norm{x^{0}-x^{*}}^{2}, 4\bar{\rho}^{2}\max\left\{\frac{r}{s},1\right\} + \frac{8r\sigma^{2}}{r-2}
    \right\}.
\end{equation*}
\end{theorem}
}

 {The proof of this result follows by a standard induction argument, and can be found in Appendix \ref{ap:convergence}. }

\section{Numerical Experiments on Electric Vehicle Charging}
\label{sec:four}
In this section, we consider a competitive game between $n$ distinct electric vehicle charging station operators, where stations are equipped with renewable power sources. The goal of each player is to set prices to maximize their own profit in a system where demand for their station will change in response to the prices set by other competing stations as well. The cost function (negative profit) takes the form
\begin{equation*}
    f_{i}(x,z_{i}) = 
    \underbrace{-z_{i}x_{i} + \frac{\lambda_{i}}{2}x_{i}^{2}}_{\text{service profit}}
    -\underbrace{p_{w}\phi(w_{i}-z_{i})}_{\text{renewable profit}}
    +\underbrace{p_{r}\phi(z_{i}-w_{i})}_{\text{operational cost}}
\end{equation*}
where $\phi(y) = \log(1+\exp(y))$ for all $y\in\real$. The renewable profit and operational cost terms allow us to describe the trade-off between profit from renewable power generation sold to the grid at rate $p_{w}$, and surplus power required from the grid to meet demand at rate $p_{r}$. 
To set prices, we can formulate a Nash equilibrium problem over the expected costs $F_{i}(x) = \bbE_{z_{i}\sim D_{i}(x)} [ f_{i}(x,z_{i})]$ for $i\in[n]$ and $x\in\cX=\Pi_{i=1}^{n} \cX_{i}$, where $\cX_{i} = [p_{w},p_{r}]$ is the interval of price values between the wholesale and retail price.

 {Since the set of reasonable prices will be quite small, we hypothesize that the the price and demand have a linear relationship of the form $z_{i} \overset{d}{=} \xi_{i} + \langle b_{i}, x_{i}\rangle $ where $b_{i}\in\real^{n}$ with $\xi_{i}\sim D_{\xi_{i}}$ corresponding to the base demand. Since we have a simple model, the first and second derivatives can be computed in closed form, and the relevant constants can be computed directly. Indeed, we find that the hypothesis of Theorem \ref{thm:sm} are satisfied with $\lambda = \min_{i}\lambda_{i}$ which we set to $1$, $L_{i} = 1$,and $\gamma_{i}=\norm{b_{i}}^{2}$. We conclude that  $G:\real^{n}\rightarrow\real^{n}$ is $\alpha = (1-2\norm{B}_{F})$-strongly monotone with  where $B$ is the parameter matrix whose columns are $b_{i}$.} 

\begin{figure}[t!]
    \centering
    \includegraphics[width=0.5\textwidth]{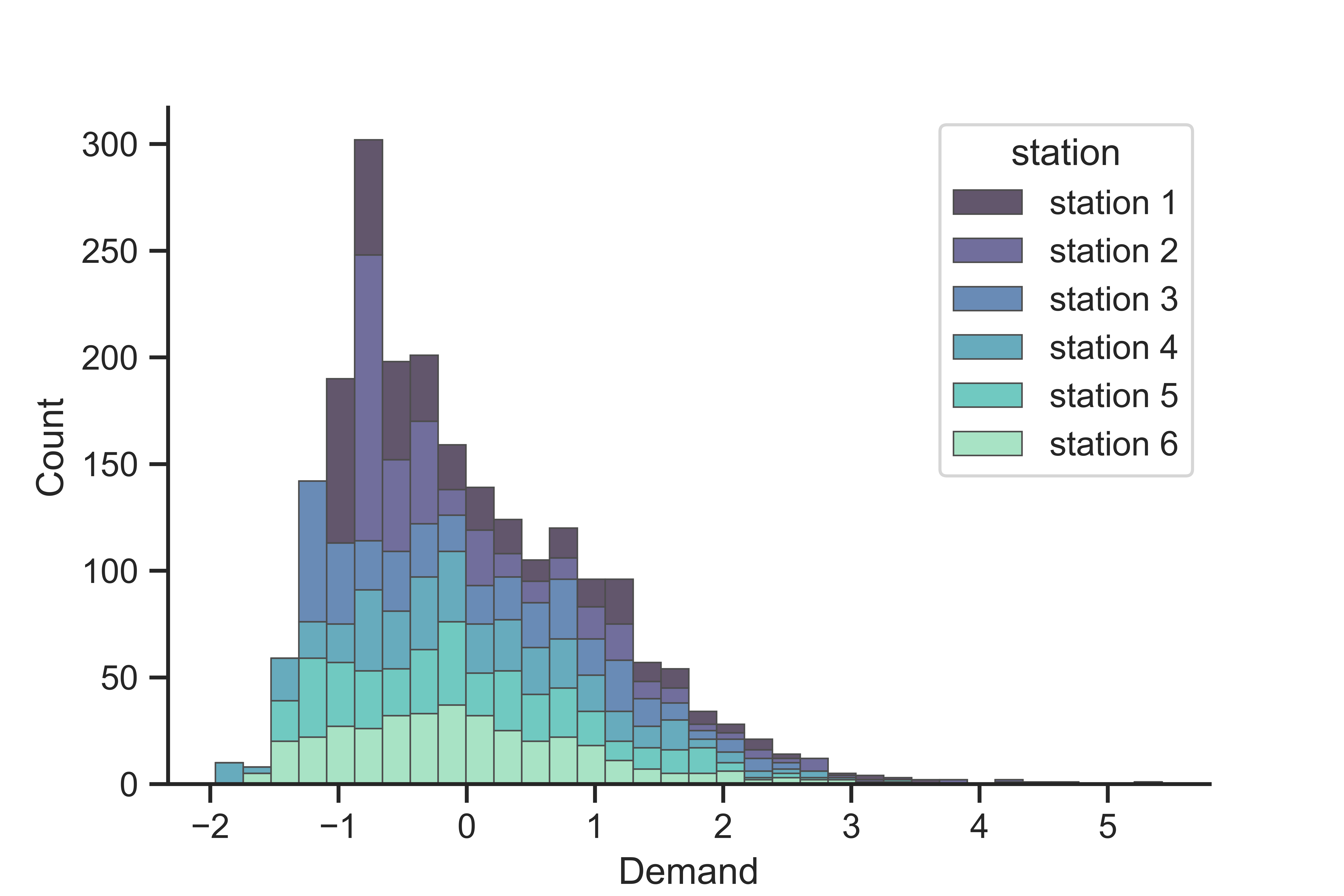}
    \caption{Standardized demand data for six medium demand EVCS's consisting of either 2 or 6 ports and port power values of 50, 150, and 350 kWh. Standardization maps raw demand instances to instances of demand that are deviations from the average at each station.}
    \label{fig:data}
\end{figure}

\begin{figure}[t!]
    \centering
    \includegraphics[width=0.5\textwidth]{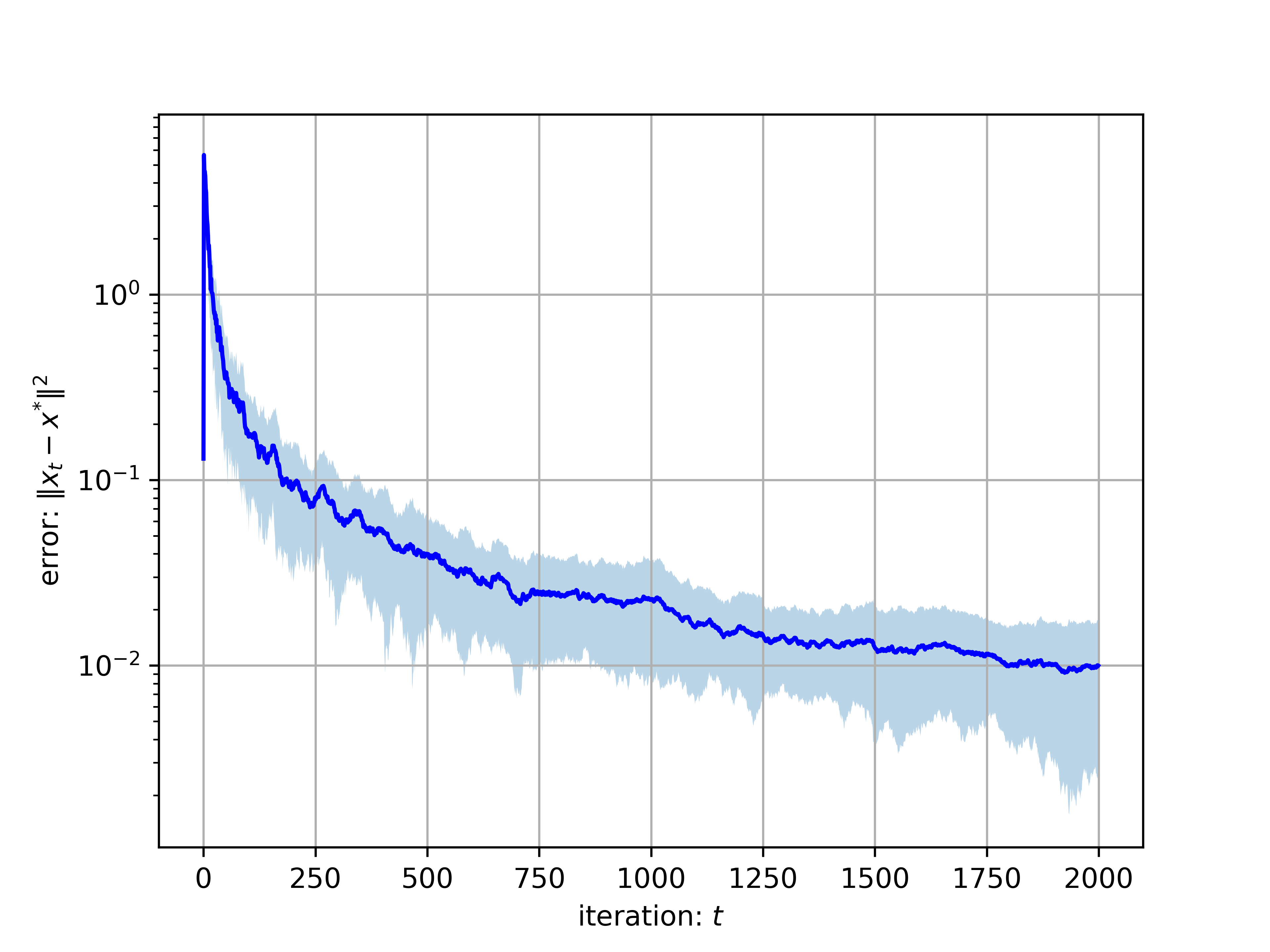}
    \caption{Expected error curve and confidence interval for regularized stochastic gradient descent with decaying step size for a location-scale model.}
    \label{fig:error}
\end{figure}

 {Our data depicts the demand of electricity across an hour-long period for 6 ports of varying power profiles for each day in year. We standardize the data to be zero mean and unit variance across each station. Solutions are calculated by performing expected gradient play with constant step size; the expected mean is estimated via the empirical mean over the data set. }

 {We set $b_{ii} = -1/18+\nu$ and $b_{ij} = 1/18+\nu$, where we use $\nu\sim\cN(0,10^{-5})$ to simulate learning $B$ from samples. Hence demand for agent $i$ decreases as their own price increases, and increases as the price of other agents decreases. We run the stochastic gradient play algorithm initialized at $x^{0} = p_{r}\bfone_{n}$ with a single sample at each round and a decaying step size policy $\omega_{t} = \alpha(r+t-2)/2$ for $r=3$. In Figure \ref{fig:error} we plot the mean error trajectory an confidence interval over 50 trials of 2000 iterations. }

\section{CONCLUSIONS}
In this work, we studied a class of stochastic Nash equilibrium problems, characterized by data distributions that are dependant on the decisions of all involved players. We showed that a learning-based approach enables the formulation of an approximate Nash equilibrium problem that is solvable using a stochastic gradient play algorithm. The results of this procedure is a cost that can be related to cost of the original Nash equilibrium problem via an error that depends on both our approximation and estimation error. To demonstrate the flexibility of these findings, we simulated these techniques in an electric vehicle charging market problem in which service providers set prices, and users modify their demand based on prices set by providers. Future research will look at a scenario where the estimate of the distributional map is improved during the operation of the algorithm, based on the feedback received. Future applications will demonstrate the efficacy of more complex models (beyond linear) 


\section*{APPENDIX}

\subsection{Proof of Lemma~\ref{lemma:gradient_bound}}
\label{ap:lemmagr}

For the sake of notation convenience, and visual clarity, we will suppress the $i$ index throughout the proof. We denote the gradient error by $J(\beta) = \nabla \hat{R}(\beta) - \nabla R(\beta)$ for all $\beta\in\real^{\ell}$.

To begin, we will generate coverings for the unit sphere in $\real^{\ell}$ and $\cB\subseteq \real^{\ell}$ and use a discretization argument to create bounds over these finite sets. Fix $\beta\in\cB$ and $u\in\cS^{\ell-1}$. Let $\{u_{j}\}_{j=1}^{N}$ be an arbitrary $1/2$-covering of the sphere $\mathbb{S}^{d_{\ell_{i}}}$ with respect to the Euclidean norm. From \cite[Lemma 5.7]{wainwright_2019}, we know that $N \leq 5^{\ell}$. From our covering, we have that there exists $u_{j}$ in the covering such that $\norm{u-u_{j}}\leq 1/2$. Hence, 
\begin{align*}
\langle u, J(\beta) \rangle 
& = \langle u_{j} + (u - u_{j}), J(\beta) \rangle \\
& = \langle u_{j}, J(\beta) \rangle + \langle u - u_{j}, J(\beta) \rangle \\
& \leq \langle u_{j}, J(\beta) \rangle + \norm{ u - u_{j} }\norm{ J(\beta) } \\ 
& \leq \langle u_{j}, J(\beta) \rangle + \frac{1}{2} \norm{ J(\beta) } \\ 
& \leq \max_{j\in[N]}\langle u_{j}, J(\beta) \rangle + \frac{1}{2} \norm{ J(\beta) } \, . 
 \end{align*}
Since this is true for any $u\in\cS^{d-1}$, then it holds for $u = J(\beta) / \norm{J(\beta)}$. Thus the above becomes 
\begin{equation}
\label{eqn:erm_intermediate}
    \norm{J(\beta)} \leq 2 \langle u_{j}, J(\beta) \rangle \leq 2\max_{j\in[N]} \langle u_{j}, J(\beta) \rangle. 
\end{equation}
Now we fix $\varepsilon\in(0,1]$, and choose and $\varepsilon$-covering for the set $\cB$, which we will write as $\{\beta_{k}\}_{k=1}^{M}$. Recall that $\cB$ is bounded, so there exists a constant $r>0$ such that for all $\beta\in\cB$, $\norm{\beta}\leq r$. Hence $\cB\subseteq B(r)$. From~\cite[Proposition 4.2.12]{vershynin2018high}, we have that 
\begin{equation}
M \leq \frac{\vol\left( B(r) + \frac{\varepsilon}{2}B(1)\right)}{\vol\left( \frac{\varepsilon}{2}B(1)\right)}  
= 
\frac{\vol\left( \frac{3}{2}B(r)\right) }{\vol\left( \frac{\varepsilon}{2}B(1)\right)}  
= \left(\frac{3r}{\varepsilon}\right)^{\ell}.
\end{equation}

Thus, we conclude that $M \leq (3r/\varepsilon)^{\ell}$.

Now by our discretization argument, there exists $k\in[M]$ such that $\norm{\beta - \beta_{k}} \leq \varepsilon$ and hence 
\begin{align*}
\max_{j\in[N]} \langle u_{j}, J(\beta) \rangle 
&  = \max_{j\in[N]} \langle u_{j}, J(\beta_{k}) + (J(\beta) - J(\beta_{k}) \rangle \\
& = \max_{j\in[N]} \langle u_{j}, J(\beta_{k}) \rangle + \langle u_{j}, J(\beta) - J(\beta_{k}) \rangle \\ 
& \leq  \max_{j\in[N]} \langle u_{j}, J(\beta_{k}) \rangle + \max_{j\in[N]} \langle u_{j}, J(\beta) - J(\beta_{k}) \rangle \\
& \leq \max_{k\in[M]} \max_{j\in[N]} \langle u_{j}, J(\beta_{k}) \rangle \\
& \ +  \sup_{\norm{\alpha - \alpha'}\leq \varepsilon} \max_{j\in[N]} \langle u_{j}, J(\alpha) - J(\alpha') \rangle. 
\end{align*}
We observe that if $\alpha,\alpha'\in\cB$ are such that $\norm{\alpha-\alpha'}\leq \varepsilon $, then applying our smoothness assumption yields
\begin{align*}
& \langle u_{j}, J(\alpha) - J(\alpha') \rangle \\
 & = \langle u_{j}, (\nabla \hat{R}(\alpha) - \nabla R(\alpha) ) - 
 (\nabla \hat{R}(\alpha') - \nabla R(\alpha') )\rangle \\
 & = \langle u_{j}, \nabla \hat{R}(\alpha) - \nabla \hat{R}(\alpha') \rangle + \langle u_{j}, \nabla R(\alpha') - \nabla R(\alpha) \rangle \\ 
 & \leq \norm{u_{j}} \norm{ \nabla \hat{R}(\alpha) - \nabla \hat{R}(\alpha')} + \norm{u_{j}} \norm{\nabla R(\alpha) - \nabla R(\alpha')} \\ 
 & \leq L_{\beta_{i}}\norm{\alpha - \alpha'} + L_{\beta_{i}}\norm{\alpha - \alpha'} \\
 & \leq 2L_{\beta}\varepsilon,
\end{align*}
where the second-to-last inequality uses  $\norm{u_{j}}=1$. 

To bound the remaining term, we use the concentration of sub-exponential random variables, due to Bernstein's Inequality combined with the Union Bound. We have that 
\begin{equation*}
    \prob\left( 
    \langle u_{j}, J(\beta_{k}) \rangle \geq t 
    \right)
    \leq 2\exp\left(-\frac{mt^{2}}{2\theta^{2}}\right)
\end{equation*}
for all $t\leq \theta$, and hence 
\begin{align*}
    & \hspace{-1.0cm} \prob\left(
    \max_{k\in[M]} \max_{j\in[N]} \langle u_{j}, J(\beta_{k} \rangle \geq t
    \right) \\ 
    & =  \prob\left(
    \bigcup_{k\in[M]}\bigcup_{j\in[N]}\{ \langle u_{j}, J(\beta_{k} \rangle \geq t \}
    \right) \\ 
    & \leq \sum_{k\in[M]} \sum_{j\in[N]} \prob\left(
    \{ \langle u_{j}, J(\beta_{k} \rangle \geq t \}
    \right) \\ 
    & \leq \sum_{k\in[M]} \sum_{j\in[N]} 2\exp\left(
    -\frac{mt^{2}}{2\theta^{2}}\right) \\
    & = M\cdot N \cdot  2\exp\left(
    -\frac{mt^{2}}{2\theta^{2}}\right) \\ 
    & \leq 2\left(\frac{15r}{\varepsilon}\right)^{\ell} \exp\left(
    -\frac{mt^{2}}{2\theta^{2}}\right)
\end{align*}
for all $t\leq \theta$, where we used the fact that $M \leq (3r/\varepsilon)^{\ell}$ and $N  \leq 5^{\ell}$. Setting the right hand side equal to $2\delta$ yields 
\begin{equation}
t = \sqrt{2}\theta\sqrt{\frac{ 
\ell\log(15r/\varepsilon) + \log(1/\delta)
}{m}}.
\end{equation}
Next we choose $\varepsilon = \frac{1}{15r}\sqrt{\frac{\ell + \log(1/\delta)}{m}}$ so that 
\begin{align*}
t & = \sqrt{2}\theta\sqrt{\frac{\ell\log(15r/\varepsilon) + \log(1/\delta)}{m}} \\ 
& = \sqrt{2}\theta\sqrt{\frac{ \frac{\ell}{2}\log(m) - \frac{\ell}{2}\log(\ell+ \log(1/\delta)) + \log(1/\delta)}{m}} \\ 
& \leq \sqrt{2}\theta\sqrt{\frac{ \ell\log(m) + \log(1/\delta)}{m}} \\ 
& \leq \sqrt{2}\theta\sqrt{\frac{ \log(m)(\ell + \log(1/\delta))}{m}}  .
\end{align*}
By requiring that $m$ satisfy $m/\log(m) \geq 2(\ell+\log(1/\delta))$, we enforce that $t\leq \theta$.
In combining, we observe that 
\begin{align*}
& t + 2\varepsilon L \\
& \leq \sqrt{2}\theta\sqrt{\frac{ \log(m)(\ell + \log(1/\delta))}{m}} 
+ \frac{2L}{15r} \sqrt{\frac{\ell + \log(1/\delta)}{m}}  \\
& \leq 2\left(\theta + \frac{L}{15r}\right)\sqrt{\frac{ \log(m)(\ell+ \log(1/\delta))}{m}} \\
& \leq 4\max\left\{ \frac{L}{15r}, \theta \right\}\sqrt{\frac{ \log(m)(\ell + \log(1/\delta))}{m}},
\end{align*}
and the result follows.

%

\subsection{Proof of Theorem~\ref{theorem:beta_estimation}}
\label{ap:beta_estimation}
We suppress the subscript $i$ for notational simplicity.  We recall that that the $\mu$-strong convexity of the map $\beta\mapsto R(x,z;\beta)$ implies $\mu$-strong monotonicity of $\nabla R(\beta)$, and $\nabla \hat{R}(\beta)$. It follows that
\begin{align*}
    \mu\norm{\hat{\beta} - \beta^{*}}^{2} & \leq \langle  \hat{\beta} - \beta^{*}, \nabla R(\hat{\beta}) -  \nabla R(\beta^{*})\rangle  \\
    & = \langle \hat{\beta} - \beta^{*}, \nabla R(\hat{\beta}) \rangle  -  \langle \hat{\beta} - \beta^{*}, \nabla R(\beta^{*})\rangle \\ 
    & \leq \langle \hat{\beta} - \beta^{*}, \nabla R(\hat{\beta}) \rangle \\
    & \leq \langle \hat{\beta} - \beta^{*}, \nabla R(\hat{\beta}) \rangle  + 
    \langle \beta^{*} - \hat{\beta} , \nabla \hat{R}(\hat{\beta}) \rangle \\
    & = \langle \hat{\beta} - \beta^{*}, \nabla R(\hat{\beta}) - \nabla \hat{R}(\hat{\beta}) \rangle \\ 
    & \leq \norm{\hat{\beta} - \beta^{*}} \sup_{\beta\in\cB} \norm{ \nabla R(\beta) - \nabla \hat{R}(\beta) }
\end{align*}
and hence 
\begin{equation}
    \norm{\hat{\beta} - \beta^{*}}\leq \frac{1}{\mu} \sup_{\beta\in\cB} \norm{ \nabla R(\beta) - \nabla \hat{R}(\beta) }.
\end{equation}
The result now follows by applying Lemma \ref{lemma:gradient_bound}.
\hfill $\Box$

\subsection{Proof of Proposition~\ref{prop:ls}}
\label{ap:propls}

We suppress the $i$ index throughout. The associated risk function is $R(x,z,B) = \frac{1}{2}\norm{Bx-z}^{2}$, so that $\nabla R(x,z,B) = (Bx-z)x^{T} = Bxx^{T} - zx^{T}$ and $\nabla^{2} R(x,z,B) = xx^{T}$ are the corresponding gradient and hessian. We observe that enforcing $\gamma I \leq \bbE[xx^{T}] \leq L I$ for some $\gamma,L>0$ ensures $\gamma$-strong convexity and $L$-smoothness of the expected risk. Similarly, the empirical risk has gradient $\nabla R_{m}(B) = 1/m(BXX^{T} - ZX^{T})$, and hessian $\nabla^{2} R_{m}(B) = (1/m)XX^{T}$. Thus $R_{m}$ is convex the hessian is symmetric, then it is positive semi-definite and thus $R_{m}$ is convex. Furthermore, smoothness of $R_{m}$ follows with constant $\max\{L,\norm{XX^{T}}_{2}\}$. Lastly, since $zx^{T}$ and $xx^{T}$ have sub-exponential entries, the gradient is sub-exponential and the result follows. 
\hfill $\Box$   
\subsection{Proof of Theorem~\ref{theorem:main}}
\label{ap:theorem_main}

We observe that for any fixed $x\in\mathcal{X}$, we have that $\vert F_{\hat{\beta}_{i}}(x) - F_{i}(x) \vert \leq  
    \vert F_{\hat{\beta}_{i}}(x) - F_{\beta_{i}^{*}}(x) \vert 
    + \vert F_{\beta_{i}^{*}}(x) - F_{i}(x) \vert$. The first term describes our statistical error at $x$. We denote $\Pi(D_{\hat{\beta}_{i}}, D_{\beta_{i}^{*}})$ as a coupling on $\cP(\real^{m_{i}})$ so that 
{\small
\begin{align*}
& \vert F_{\hat{\beta}_{i}}(x) - F_{\beta_{i}^{*}}(x) \vert \\
& = \left\vert \inf_{\Pi(D_{\hat{\beta}_{i}}(x), D_{\beta_{i}^{*}}(x))} \ \bbE_{(z,z')\sim \Pi(D_{\hat{\beta}_{i}}(x), D_{\beta_{i}^{*}}(x))} \left(f(x,z) - f(x,z')\right) \right\vert \\
& \leq \inf_{\Pi(D_{\hat{\beta}_{i}}(x), D_{\beta_{i}^{*}}(x))} \ \bbE_{(z,z')\sim \Pi(D_{\hat{\beta}_{i}}(x), D_{\beta_{i}^{*}}(x))} \left\vert f(x,z) - f(x,z')\right\vert \\
& \leq L_{z_{i}} \left(\inf_{\Pi(D_{\hat{\beta}_{i}}(x), D_{\beta_{i}^{*}}(x))} \ \bbE_{(z,z')\sim \Pi(D_{\hat{\beta_{i}}}(x), D_{\beta_{i}^{*}}(x))} \norm{z_{i} - z_{i}^{'}}\right) \\ 
& = L_{z_i}W_{1}(D_{\hat{\beta}_{i}}(x), D_{\beta_{i}^{*}}(x)) \\ 
& \leq L_{z_{i}} \varepsilon_{i} \norm{ \hat{\beta}_{i} -  \beta_{i}^{*}} \, .
\end{align*}
}
By similar argument, we find that $\vert F_{\beta_{i}^{*}}(x) - F_{i}(x) \vert 
    \leq L_{z_{i}} W_{1}( D_{\beta_{i}^{*}}(x) , D_{i}(x) ) 
    \leq L_{z_{i}} \gamma_{i}$.
In combining, we get $\vert F_{\hat{\beta}_{i}}(x) - F_{i}(x) \vert 
     \leq  L_{z_{i}} \varepsilon_{i} \norm{ \hat{\beta}_{i} -  \beta_{i}^{*}}  + L_{z_{i}} \gamma_{i}$. Lastly, $\norm{ \hat{\beta}_{i} -  \beta_{i}^{*}}$ can be bounded as in Theorem~\ref{theorem:beta_estimation}. 
     
Regarding the second bound, we have that 
\begin{align*}
    & F_{i}(\hat{x}) - F_{i}(x^{*}) \\
    & = \left[ F_{i}(\hat{x}) - F_{\beta_i^{*}}(\hat{x}) \right] 
    + \left[ F_{\beta_i^{*}}(\hat{x}) - F_{\hat{\beta_i}}(\hat{x}) \right] \\
    & \ + \left[ F_{\hat{\beta_i}}(\hat{x}) - F_{\hat{\beta_i}}(x^{**})\right] 
     + \left[ F_{\hat{\beta_i}}(x^{**}) - F_{\beta_i^{*}}(x^{**}) \right] \\
    & \ + \left[ F_{\beta_i^{*}}(x^{**}) - F_{\beta_i^{*}}(x^{*})\right] 
    + \left[ F_{\beta_i^{*}}(x^{*}) - F_{i}(x^{*}) \right] \\
    & \leq 2 \norm{ F_{i} - F_{\beta_i^{*}}}_{\infty} + 2 \norm{ F_{\beta_i^{*}} - F_{\hat{\beta_i}}}_{\infty} \\
    & \ + \left[ F_{\hat{\beta_i}}(\hat{x}) - F_{\hat{\beta_i}}(x^{**})\right] + \left[ F_{\beta_i^{*}}(x^{**}) - F_{\beta_i^{*}}(x^{*})\right] 
\end{align*}
where $x_i^{**} \in \mathcal{X}^{**}$, where $\mathcal{X}^{**}$ is the set of equilibria of
\begin{equation}
    {x}_{i}^{**} \in \argmin_{x_{i}\in\cX_i} F_{{\beta}^*_{i}}(x_{i},{x}^{**}_{-i}), \quad i \in [n]
\end{equation}
where 
$$
F_{{\beta}^*_{i}}(x_{i},{x}^{**}_{-i}) := \hspace{-.2cm}
    \underset{z_{i}\sim D_{{\beta}^*_{i}}(x_{i},{x}^{**}_{-i})}{\bbE}
    f_{i}(x_{i},{x}^{**}_{-i},z_{i}) \, .
$$
Then,
\begin{align*}
F_{\hat{\beta_i}}(\hat{x}) - F_{\hat{\beta_i}}(x^{**}) &  \leq \left[ F_{\hat{\beta_i}}( \hat{x}_{i}, \hat{x}_{-i}) - F_{\hat{\beta_i}}(x^{**}_{i},  \hat{x}_{-i}) \right] \\
& \ + \left[ F_{\hat{\beta_i}}( x^{**}_{i}, \hat{x}_{-i})  - F_{\hat{\beta_i}}(x^{**}_{i}, x^{**}_{-i}) \right] \\
& \leq  L_{i}^{\hat{\beta}} \|\hat{x}_i - x_i^{**}\| +  L_{-i}^{\hat{\beta}} \norm{ \hat{x}_{-i} -  x^{**}_{-i}} \\
& \leq \sqrt{2} \max\{L_{i}^{\hat{\beta}},L_{-i}^{\hat{\beta}}\} \norm{ \hat{x} -  x^{**}}
\end{align*}
where we have used the inequality $\sqrt{a} + \sqrt{b} \leq \sqrt{2} \sqrt{a+b}$ for some $a, b \geq 0$. 

Next, consider
\begin{align*}
     F_{\beta^{*}_i}(x^{**}) - F_{\beta^{*}_i}(x^{*}) &  = [F_{\beta^{*}_i}(x_i^{**},x_{-i}^{**})- F_{\beta^{*}_i}(x_i^{**},x^{*}_{-i})] \\
    & ~~ + [F_{\beta^{*}_i}(x_i^{**},x^{*}_{-i}) - F_{\beta^{*}_i}(x_i^{*}, x_{-i}^{*})] \\
    & \leq L_{-i}^{\beta^*} \|x_{-i}^{**} - x_{-i}^{*}\| + L_{i}^{\beta^*} \|x_{i}^{**} - x_{i}^{*}\| \\
    & \leq \sqrt{2} \max\{L_{i}^{\beta^*}, L_{-i}^{\beta^*}\} \|x^{**} - x^{*}\| \, .
\end{align*}
Combining the bounds yields 
\begin{align*}
        & F_{i}(\hat{x}) - F_{i}(x^{*})  \\
        & \leq 2 \norm{ F_{i} - F_{\beta^{*}_i}}_{\infty} + 2 \norm{ F_{\beta^{*}_i} - F_{\hat{\beta_i}}}_{\infty} \\
        & ~~ +  \sqrt{2} \max\{L_{i}^{\hat{\beta}},L_{-i}^{\hat{\beta}}\} \norm{ \hat{x} -  x^{**}} \\
        & ~~ + \sqrt{2} \max\{L_{i}^{\beta^*}, L_{-i}^{\beta^*}\} \|x^{**} - x^{*}\| \\
        & \leq 2\gamma_{i} L_{z_{i}} + 2\varepsilon L_{z_{i}} \norm{ \hat{\beta}_{i} -  \beta_{i}^{*}} \\
        & ~~ + \sqrt{2}(\max\{L_{i}^{\hat{\beta}},L_{-i}^{\hat{\beta}}\} + \max\{L_{i}^{\beta^*}, L_{-i}^{\beta^*}\}) \diam(\mathcal{X})\\
        & \leq 2\gamma_{i} L_{z_{i}} + 2\varepsilon L_{z_{i}} \norm{ \hat{\beta}_{i} -  \beta_{i}^{*}} + 2 \sqrt{2} \bar{L}_i \diam(\mathcal{X})
\end{align*}
Then, \eqref{eqn:excess_risk} follows using the bound on $\norm{\hat{\beta}_i - \beta_i^*}$ from Theorem~\ref{theorem:beta_estimation}.
\hfill $\Box$

\subsection{Proof of Lemma~\ref{lem:step_improvement}}
\label{ap:step_improvement}
 {Consider the function $\varphi:\real^{d}\rightarrow \real$ defined by $\varphi(y) = \frac{1}{2}\norm{x^{t}-W^{-1}_{i}g_{i}^{t} -y}_{W}^{2}$ for all $y\in\cX$. Then, $\varphi$ is $\omega_{n}$-strongly convex over $\cX$ and has a unique minimizer $x^{t+1}\in\cX$. This implies that:  
\begin{equation*}
    \varphi(x^{*}) \geq \varphi(x^{t+1}) + \langle x^{*}-x^{t+1} , \nabla \varphi(x^{t+1}) \rangle  +\frac{\omega_{n}}{2}\norm{x^{t+1}-x^{*}}^{2} \, .
\end{equation*}
Since $\langle x- x^{t+1}, \nabla \varphi(x^{t+1}) \rangle \geq 0$ for all $x\in\cX$, we obtain
\begin{equation*}
    \norm{x_{i}^{k+1} - x_{i}^{*}} \leq \norm{x_{i}^{k}-\eta_{i}g_{i}^{k} - x_{i}^{*}}_{W}^{2} - \norm{x_{i}^{k}-\eta_{i}g_{i}^{k} - x_{i}^{k+1}}_{W}^{2}.
\end{equation*}
It follows that 
\begin{align*}
    \frac{\omega_{1}}{\omega_{n}}\norm{x^{t+1} - x^{*}}_{W}^{2} 
    & \leq \norm{x^{t}-x^{*}}_{W}^{2} - \norm{x^{t}-x_{i}^{t+1}}_{W}^{2} \\
    & - 2\langle x^{t} - x^{*}, g^{t} \rangle + 2\eta_{i}\langle x^{t} - x^{t+1}, g^{t} \rangle .
\end{align*}
We now consider the above in the conditional expectation $\bbE_{t} \cdot :=\bbE_{z_{i}\sim D(x_{t})}[\ \cdot \ \vert\cF_{t}]$ with $\cF_{t}=\sigma(g^{t},\tau\geq t)$. We find that
\begin{align*}
     & \frac{\omega_{1}}{\omega_{n}}\bbE_{t}\norm{x^{t+1} - x^{*}}_{W}^{2} \\
     & \leq \bbE_{t}\norm{x^{t}-x^{*}}_{W}^{2} - \bbE_{t}\norm{x^{t}-x^{t+1}}_{W}^{2} \\
     &  ~~ - 2\bbE_{t}\langle x^{t} - x^{*}, g_{i}^{t} \rangle - 2\bbE_{t}\langle x^{t+1} - x^{t}, g^{t} \rangle \\ 
     & = \norm{x^{t}-x^{*}}_{W}^{2} - \bbE_{t}\norm{x^{t}-x^{t+1}}_{W}^{2}\\
     & ~~ - 2\langle x^{t} - x^{*}, \mu^{t} \rangle - 2\bbE_{t}\langle x^{t+1} - x^{t}, g^{t} \rangle  \\
     & = \norm{x^{t}-x^{*}}_{W}^{2} - \bbE_{t}\norm{x^{t}-x^{t+1}}_{W}^{2} \\
     & ~~ + 2\bbE_{t}\langle x^{t}-x^{t+1}, g^{t} - \mu^{t}\rangle + 2\bbE_{t}\langle x^{*}-x^{t+1}, \mu^{t}\rangle \\
    & = \norm{x^{t}-x^{*}}_{W}^{2} - \bbE_{t}\norm{x^{t}-x^{t+1}}_{W}^{2} - 2\langle x^{t+1}-x^{*}, G(x^{t+1})\rangle\\
    & ~~ + 2\bbE_{t}\langle x^{*}-x^{t+1}, \mu^{t} - G(x^{t+1})\rangle \\
    & ~~ + 2\bbE_{t}\langle x^{t}-x^{t+1}, g^{t} - \mu^{t}\rangle.
\end{align*}
To proceed, we bound the inner product terms. Using strong monotonicity, we have that
\begin{align*}
    \bbE_{t} \langle x^{*} - x^{t+1}, G(x^{t+1}) \rangle  
    & \geq \alpha \bbE_{t}\norm{x^{t+1}-x^{*}}^{2} \\
    & \geq \frac{\alpha}{\omega_{1}} \bbE_{t} \norm{x^{t+1}-x^{*}}_{W}^{2}. 
\end{align*}
Furthermore, we observe that 
\begin{align*}
     & \bbE_{t}\langle x^{*}-x_{i}^{t+1}, \mu^{t} - G(x^{t+1})\rangle \\
     & \hspace{1.0cm} =  \bbE_{t}\langle x^{*}-x^{t+1}, \mu^{t} - G(x^{t})\rangle  \\
     & \hspace{1.0cm} + \bbE_{t}\langle x^{*}-x^{t+1}, G(x^{t}) - G(x^{t+1})\rangle  .
\end{align*}
To bound the remaining terms, we use arguments based on a weighted Young's inequality. Let $\Delta_{1}, \Delta_{2},\Delta_{3}>0$ be fixed constants. It follows that 
\begin{align*}
    & 2\bbE_{t}\langle x^{t} - x^{t+1}, g^{t} - \mu^{t}\rangle \\
    & \leq \Delta_{1} \bbE_{t} \norm{x^{t+1} - x^{t}}^{2} + \frac{1}{\Delta_{1}}\bbE_{t}\norm{g^{t} - \mu^{t}}^{2} \\ 
    & \leq \frac{\Delta_{1}}{\omega_{n}} \bbE_{t} \norm{x^{t+1} - x^{t}}_{W}^{2} + \frac{1}{\Delta_{1}}\sum_{i=1}^{n} \bbE_{t}\norm{g^{t} - \mu^{t}}^{2} \\ 
    & \leq \frac{\Delta_{1}}{\omega_{n}} \bbE_{t} \norm{x^{t+1} - x^{t}}_{W}^{2} + \frac{1}{\Delta_{1}}\sum_{i=1}^{n} \sigma_{i}^{2} \\
    & \leq \frac{\Delta_{1}}{\omega_{n}} \bbE_{t} \norm{x^{t+1} - x^{t}}_{W}^{2} + \frac{\sigma^{2}}{\Delta_{1}}, \\
\end{align*}
and
\begin{align*}
    & 2\bbE_{t}\langle x^{*} - x^{t+1}, \mu^{t} - G(x^{t}) \rangle \\
    & \leq \Delta_{2} \bbE_{t} \norm{x^{t+1} - x^{*}}^{2} + \frac{1}{\Delta_{2}}\bbE_{t}\norm{\mu^{t} - G(x^{t})}^{2} \\ 
    & \leq \frac{\Delta_{2}}{\omega_{n}} \bbE_{t} \norm{x^{t+1} - x^{*}}_{W}^{2} + \frac{1}{\Delta_{2}}\sum_{i=1}^{n} \bbE_{t}\norm{\mu^{t} - G(x^{t})}^{2} \\ 
    & \leq \frac{\Delta_{2}}{\omega_{n}} \bbE_{t} \norm{x^{t+1} - x^{*}}_{W}^{2} + \frac{1}{\Delta_{2}}\sum_{i=1}^{n} \rho_{i}^{2} \\
    & \leq \frac{\Delta_{2}}{\omega_{n}} \bbE_{t} \norm{x^{t+1} - x^{*}}_{W}^{2} + \frac{\rho^{2}}{\Delta_{2}}. \\
\end{align*}
Additionally, we have that
\begin{align*}
    & 2\bbE_{t}\langle x^{*} - x^{t+1}, G(x^{t}) - G(x^{t+1}) \rangle \\  
    & \leq \Delta_{3} \bbE_{t} \norm{x^{t+1} - x^{*}}^{2} + \frac{1}{\Delta_{3}}\bbE_{t}\norm{G(x^{t}) - G(x^{t+1})}^{2} \\ 
    & \leq \frac{\Delta_{3}}{\omega_{n}} \bbE_{t} \norm{x^{t+1} - x^{*}}_{W}^{2} + \frac{L^{2}}{\Delta_{3}} \bbE_{t}\norm{x^{t+1} - x^{t}}^{2} \\ 
    & \leq \frac{\Delta_{3}}{\omega_{n}} \bbE_{t} \norm{x^{t+1} - x^{*}}_{W}^{2} + \frac{L^{2}}{\omega_{n}\Delta_{3}} \bbE_{t}\norm{x^{t+1} - x^{t}}^{2}_{W}. \\ 
\end{align*}
Combining these estimates yields 
\begin{align*}
& \frac{\omega_{n}}{\omega_{1}}\bbE_{t}\norm{x^{t+1}-x^{*}}^{2}_{W} \\ 
& \leq  \norm{x^{t} - x^{*}}_{W}^{2} - \bbE_{t} \norm{x^{t+1}-x^{t}}_{W}^{2} 
- \frac{2\alpha}{\omega_{1}} \bbE_{t} \norm{x^{t+1}-x^{*}}_{W}^{2}  \\
& ~~+\left(\frac{\Delta_{1}}{\omega_{n}} \bbE_{t} \norm{x^{t+1} - x^{t}}_{W}^{2} + \frac{\sigma^{2}}{\Delta_{1}} \right) \\
& ~~ + \left( \frac{\Delta_{2}}{\omega_{n}} \bbE_{t} \norm{x^{t+1} - x^{*}}_{W}^{2} + \frac{\rho^{2}}{\Delta_{2}} \right) \\ 
& ~~ + \left( \frac{\Delta_{3}}{\omega_{n}} \bbE_{t} \norm{x^{t+1} - x^{*}}_{W}^{2} + \frac{L^{2}}{\omega_{n}\Delta_{3}} \bbE_{t}\norm{x^{t+1} - x^{t}}^{2}_{W} \right) \\ 
& = \norm{x^{t} - x^{*}}_{W}^{2} + \left(\frac{\Delta_{1}}{\omega_{n}} + \frac{L^{2}}{\omega_{n}\Delta_{3}}-1\right)\bbE_{t} \norm{x^{t+1}-x^{t}}_{W}^{2} \\
& ~~+ \left( \frac{\Delta_{2}}{\omega_{n}} + \frac{\Delta_{3}}{\omega_{n}} - \frac{2\alpha}{\omega_{1}}\right)\bbE_{t} \norm{x^{t+1}-x^{*}}_{W}^{2}  \\ 
& ~~+ \left( \frac{\sigma^{2}}{\Delta_{1}} + \frac{\rho^{2}}{\Delta_{2}}
\right) 
\end{align*} 
and simplifying gives 
\begin{align*}
    & \left(\frac{\omega_{n}}{\omega_{1}} + \frac{2\alpha}{\omega_{1}} - \frac{\Delta_{2}}{\omega_{n}} - \frac{\Delta_{3}} {\omega_{n}}\right)\bbE_{t}\norm{x^{t+1} - x^{*}}_{W}^{2}\\
    & \leq \norm{x^{t} - x^{*}}_{W}^{2} +  \left( \frac{\sigma^{2}}{\Delta_{1}} + \frac{\rho^{2}}{\Delta_{2}}
    \right) \\
    & ~~ + \left(\frac{\Delta_{1}}{\omega_{n}} + \frac{L^{2}}{\omega_{n}\Delta_{3}}-1\right)\bbE_{t} \norm{x^{t+1}-x^{t}}_{W}^{2} . 
\end{align*}
To proceed, we choose $\Delta_{2}=\Delta_{3} = \frac{\alpha\omega_{1}}{\omega_{n}}$ and $\Delta_{1}=\omega_{1}^{-1} - 2\omega_{1}L^{2}/(\alpha\omega_{n})$ to ensure that the coefficient on the $\bbE_{t}\norm{x^{t+1}-x^{t}}_{W}^{2}$ term is zero. Furthermore, enforcing that $\frac{\omega_{1}}{\omega_{n}^{2}}\leq \frac{\alpha}{4L^{2}}$ guarantees that $\Delta_{1}^{-1} \leq 2\omega_{n}^{-1}$. Hence the variance term is finite. Substituting these values and simplifying yields the result}.
\hfill $\Box$   

\subsection{Proof of Theorem \ref{thm:nbhd_convergence}}
\label{ap:nbhd_convergence}
 {For notational convenience, we will use the short-hand notation $e^{t} := \norm{x^{t}-x^{*}}_{W}^{2}$, $c = \omega_{1} / (\alpha + \omega_{n}) $, and 
\begin{equation*}
    A = 2\frac{ \alpha\sigma^{2} + \omega_{1}\rho^{2}}{\alpha\omega_{n}} \, .
\end{equation*}
Hence, the result in Lemma \ref{lem:step_improvement} can be written compactly as 
\begin{equation*}
\bbE_{t-1}e^{t}\leq ce^{t-1} + cA.
\end{equation*}
By recursively applying this result and applying the law of total expectation, we find that
\begin{align*}
    \bbE e^{t}  \leq c^{t} e^{0} + cA\sum_{j=1}^{t-1} c^{j} \leq c^{t}e^{0} + cA\frac{1-c^{t}}{1-c}. \\ 
\end{align*}
Furthermore, if $(\omega_{1} - \omega_{n}) < \alpha$, then $c <1$ and the geometric series converges and is equal to its limit $1/(1-c)$. Hence $\bbE e^{t} \leq c^{t}e^{0} + A\frac{c}{1-c}$}. 
\hfill $\Box$   

\subsection{Proof of Theorem \ref{thm:convergence}}
\label{ap:convergence}
 {Fix $t\geq0$. For notational convenience, we will denote $e^{t} = \norm{x^{t}-x^{*}}^{2}$. Replacing the step-size matrix in Lemma \ref{lem:step_improvement} with $W=\omega^{t}I_{d\times d}$ yields
\begin{equation}
\label{eqn:deacying_step_result}
\bbE_{t}e^{t+1} \leq \frac{\omega^{t}}{\omega^{t} + \alpha}e^{t} + \frac{2\sigma^{2}}{\omega^{t}(\omega^{t} + \alpha)} + \frac{2(\rho^{t})^{2}}{\alpha(\omega^{t} + \alpha)}.
\end{equation}
To proceed, we will use the observation that 
\begin{equation}
    \frac{1}{(s+t)(r+t)} = \frac{r+t}{(s+t)(r+t)^{2}} \leq \frac{\max\{\frac{r}{s},1\}}{(r+t)^2}
\end{equation}
and
\begin{equation}
    \frac{1}{(r+t)(r+t-2)} \leq \frac{\frac{r}{r-2}}{(r+t)^2}. 
\end{equation}
By substituting our expression for $\omega^{t}$, $\rho^{t}$, and $e^{t}$ into \eqref{eqn:deacying_step_result} we obtain 
\begin{align*}
    \bbE_{t}e^{t+1} 
    & \leq \frac{r+t-2}{\alpha^{2}(r+t)^{2}}A + \frac{8\sigma^{2}}{\alpha^{2}(r+t-2)(r+t)} \\
    & ~~ + \frac{4\bar{\rho}}{\alpha^{2}(s+t)(r+t)} \\ 
    & \leq \frac{r+t-2}{\alpha^{2}(r+t)^{2}}A + \frac{8\sigma^{2}\left(\frac{r}{r-2}\right)}{\alpha^{2}(r+t)^{2}} + \frac{4\bar{\rho}\max \left\{ \frac{r}{s},1 \right\}}{\alpha^{2}(r+t)^{2}} \\
    & = \frac{r+t-1}{\alpha^{2}(r+t)^{2}}A \\
    & ~~ + \frac{-A + 8\sigma^{2}\left(\frac{r}{r-2}\right)+ 4\bar{\rho}\max \left\{ \frac{r}{s},1 \right\}}{\alpha^{2}(r+t)^{2}}  \\
    & \leq \frac{r+t-1}{\alpha^{2}(r+t)^{2}}A \\ 
    & \leq \frac{A}{\alpha^{2}(r+t+1)}.
\end{align*}
Here, the last steps follow from construction of $A$, and the fact that $(r+t+1)(r+t-1)\leq (r+t)^{2}$}.
\hfill $\Box$   

\section*{ACKNOWLEDGMENTS}
This research was supported in part by the National Science Foundation (NSF) Mathematical Sciences Graduate Internship (MSGI) Program and by NSF awards 1941896 and 2044946.  

The MSGI program is administered by the Oak Ridge Institute for Science and Education (ORISE) through an inter-agency agreement between the U.S. Department of Energy (DOE) and NSF. ORISE is managed for DOE by ORAU. All opinions expressed in this paper are the author's and do not necessarily reflect the policies and views of NSF, ORAU/ORISE, or DOE.

\bibliographystyle{IEEEtran}
\bibliography{references.bib}

\end{document}